\documentclass[superscriptaddress,amsmath,amssymb,prd,preprintnumbers,showpacs,twocolumn,nofootinbib]{revtex4-2}
\usepackage[colorlinks=true,pdfstartview=FitV,linkcolor=blue,citecolor=blue,urlcolor=blue,breaklinks=true]{hyperref}
\usepackage{graphicx}
\usepackage[T1]{fontenc} 
\usepackage{float} 
\usepackage{amssymb,amsmath,bm,natbib}
\usepackage{color}
\usepackage{slashed}
\usepackage{graphics}
\usepackage{graphicx}
\usepackage[utf8]{inputenc}
\usepackage[caption=false]{subfig}
\usepackage{hyperref}
\usepackage{url}
\usepackage{dsfont}
\usepackage{float}
\usepackage{cancel}
\usepackage{units}
\usepackage{blindtext}
\usepackage[utf8]{inputenc}
\usepackage{upgreek}
\usepackage{booktabs}
\usepackage[dvipsnames,table,xcdraw]{xcolor}
\usepackage{enumerate}
\usepackage{mathtools}
\usepackage{soul,color}
\usepackage{amsmath}
\usepackage{amssymb}
\usepackage[normalem]{ulem}
\newcommand{\orcid}[1]{\href{https://orcid.org/#1}{\includegraphics[width=10pt]{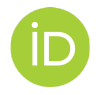}}}
\usepackage[T1]{fontenc} 
\usepackage{xcolor}
\usepackage{graphicx}
\usepackage{cleveref}
\usepackage{amsopn}
\usepackage{braket}
\usepackage{bbold}

\def\m{\mu}

\def \be {\begin{equation}}
\def \ee {\end{equation}}
\def \bea {\begin{eqnarray}}
\def \eea {\end{eqnarray}}

\def \be {\begin{equation}}
\def \ee {\end{equation}}
\def \bea {\begin{eqnarray}}
\def \eea {\end{eqnarray}}

\begin{document}
\title{Dipole ghosts and spontaneous symmetry breaking in higher-order Chern-Simons theory}
 \author{Carlos M. Reyes\orcid{0000-0001-5140-6658}}
\email{creyes@ubiobio.cl}
\affiliation{Centro de Ciencias Exactas, Facultad de Ciencias, 
Universidad del B\'{i}o-B\'{i}o, Chill\'{a}n, Chile}
\author{C\'esar Riquelme\orcid{0000-0003-0837-3891}}
\email{cesar.riquelme@uss.cl}
\affiliation{Facultad de Ingenier\'{i}a, Universidad San Sebasti\'{a}n, Lientur 1457, Concepci\'{o}n, Chile}
  \author{Angel Sanchez\orcid{0000-0002-8237-5257}}
\email{ ansac@ciencias.unam.mx }
\affiliation{Departamento de F\'isica, Facultad de Ciencias, 
Universidad Nacional Aut\'onoma de M\'exico,
Apartado Postal 70-542, Ciudad de M\'exico 04510, Mexico}
\begin{abstract}
In this work, we investigate the emergence of dipole ghost 
structures in gauge theories at both the classical and quantum levels. Traditionally, dipole 
ghosts are introduced through an auxiliary field $\chi$ satisfying $\Box^2\chi=0$, whose 
presence is reflected in the appearance of double poles in the propagator. We show 
that such dipole ghost sectors can instead be understood as a consequence of the 
multiplicity of solutions of the classical equations of motion. To establish this 
connection, we develop a constructive method and first apply it to covariant 
Maxwell theory in $(2+1)$ dimensions, where the essential ingredients can be identified 
in a transparent manner. We then extend the analysis to the constrained higher-order 
Chern-Simons theory, demonstrating that the same mechanism gives rise to dipole 
ghost sectors and associated double poles in the propagator. Our results provide a 
unified perspective on the origin of dipole ghosts, relating them directly to the 
degeneracy structure of the underlying classical dynamics. As a physical application, 
we investigate spontaneous symmetry breaking 
and compute the one-loop effective potential. We show that symmetry breaking 
removes the degeneracy underlying the dipole ghost sector, leading to a 
spectrum of ordinary massive gauge excitations that determine the quantum vacuum structure of the theory.
\end{abstract}
\maketitle
\section{Introduction}
Chern-Simons (CS)
gauge theories have been extensively studied over the past 
decades due to their rich mathematical structure and wide range of 
physical applications. In $(2+1)$ dimensions, the CS term generates 
a topological mass for the gauge field while preserving gauge invariance~\cite{Schonfeld:1980kb,Deser:1981wh}. 
Moreover, CS models have found important applications in condensed-matter 
physics, particularly in the description of the quantum Hall 
effect~\cite{Girvin:1987fp,Zhang:1988wy,Hansson:2016zlh}, anyonic 
excitations~\cite{Iengo:1991zbc} and topological 
phases of matter~\cite{Hasan:2010xy,vonDossow:2025bwr}. Also in the study of 
spontaneous symmetry breaking~\cite{Kondo:1994cz,Irvine:2001uv},
radiative corrections~\cite{Chen,C-H,Bashir:2008ej,Concha-Sanchez:2013wre,Belich:2002jd},
classical approaches~\cite{Belich:2004pv} and many other contexts, see the review~\cite{Dunne:1998qy}.

Motivated by these applications, several extensions of CS 
electrodynamics have been proposed. Among them, higher-derivative generalizations 
have attracted particular attention because they modify the ultraviolet behavior of the theory while 
preserving many of its topological features. Such extensions 
naturally arise from the effective action~\cite{JTD} and have been applied in
Lee-Wick-type constructions to study unitarity~\cite{Reyes:2009zb,Avila:2019xdn,Avila:2025jxv}
and Proca extensions~\cite{Sararu:2014sua,Sararu:2016qnx}.
The presence of higher derivatives enriches the spectrum of the theory and may
lead to new dynamical phenomena absent in ordinary Maxwell-Chern-Simons electrodynamics.



Dipole ghost structures are known to arise in covariant quantization schemes, 
where generalized solutions are associated with higher-order differential operators~\cite{Mikhov:1981fu,Karowski:1974qx,Yokoyama:1975aw,Narnhofer:1978sw}. 
In covariant Maxwell theory, for example, the longitudinal sector contains a 
hidden dipole structure whose presence is reflected in the appearance of a double pole in the propagator. More generally, 
dipole ghosts are frequently introduced through auxiliary fields 
satisfying fourth-order equations of motion, while the 
corresponding double pole is regarded as their defining signature.
Critical theories of gravity provide a notable example, where the linearized field equations develop degenerate 
fourth-order operators of the form $(\Box+\frac{2 \Lambda}{3})^2$, leading to logarithmic generalized modes~\cite{Lu:2011zk,Grumiller:2008qz}. Also, the
singleton dipole~\cite{Flato:1986uh} and cosmological perturbations 
around special backgrounds~\cite{Myung:2014pza}. At the quantum level, 
they may give rise to unconventional sectors involving states of indefinite norm and dipole ghost degrees of freedom~\cite{Nakanishi:1966zz}. 
Understanding the origin and physical significance of these structures remains an important problem in the quantization of gauge theories.

Despite their long history, the dynamical origin of dipole ghost 
sectors is often obscured. In particular, the connection 
between repeated roots of the classical equations of motion, generalized 
mode solutions, Jordan chains, mixed oscillator 
algebras, and the emergence of negative-norm states is 
rarely made explicit. It is therefore natural to ask whether 
dipole ghosts should be viewed as fundamental ingredients of the 
theory or rather as manifestations of an underlying degeneracy in the classical dynamics.

Degenerate sectors associated with repeated roots of the characteristic polynomial arise in a variety of physical 
contexts. Examples range from critically damped systems~\cite{Goldstein}
to $S$-matrix physics and decaying processes~\cite{Hernandez:2002otf,Bohm:1997mr}.
Such degeneracies are closely related to non-diagonalizable 
evolution operators and Jordan-block structures.

In this work, we address these questions in the context of higher-order CS 
electrodynamics in $(2+1)$ dimensions. We develop a constructive framework that 
relates the multiplicity of classical solutions to the emergence of dipole ghost 
sectors in the quantum theory. To establish this connection, we first analyze 
covariant Maxwell theory, where the essential ingredients can be identified 
in a particularly transparent way. 
We then extend the analysis to the higher-order CS model, where 
the same mechanism is shown to operate in the presence of constraints.
By explicitly solving the equations of motion in momentum space, 
we show that the characteristic polynomial develops roots with multiplicity two. 
As a consequence, the solutions are not exhausted by ordinary plane waves but 
instead contain generalized modes proportional to $t e^{-i\omega t}$. These modes 
signal the presence of a non-diagonalizable dynamical sector associated with a 
Jordan-block structure. The explicit resolution of the field equations 
allows us to construct the corresponding mode expansion and identify the independent 
sectors of the theory. We show that the degenerate sector generates a mixed oscillator 
algebra rather than a collection of independent canonical oscillators. After 
diagonalization of operators, positive, and negative-norm modes emerge naturally, 
making the dipole ghost structure manifest. To our knowledge, this is the first constructive derivation in which dipole ghost sectors emerge directly from the multiplicity of classical solutions, without introducing auxiliary fourth-order fields.

As an application, we investigate the effects of spontaneous 
symmetry breaking in the higher-order CS model. 
We show that the Higgs mechanism lifts the degeneracy responsible 
for the dipole ghost sector, replacing it with a spectrum 
of ordinary massive gauge excitations. The corresponding one-loop 
effective potential is computed and used to analyze the resulting vacuum structure.

The paper is organized as follows. In Sec.~\ref{sectionII}, we develop the method in the context 
of covariant Maxwell theory in $(2+1)$ dimensions. We solve the equations of motion, identify 
the degenerate sector and the associated generalized modes, and construct 
the corresponding quantum theory. In Sec.~\ref{SectionIII}, we extend the analysis to the 
higher-order CS theory, where we identify the dipole ghost 
sector and derive the associated Feynman propagator. In Sec.~\ref{SSBIV}, we investigate the 
effects of spontaneous symmetry breaking by coupling the gauge 
sector to a charged scalar field. We show that 
the resulting gauge-boson mass lifts the degeneracy responsible 
for the dipole ghost sector and compute the corresponding 
one-loop effective potential. Finally, in 
Sec.~\ref{sec:conclusions}, we summarize our results and discuss their 
implications for the quantization of theories 
with degenerate dynamical sectors. In Appendix~\ref{App:A},
we derive the creation and annihilation operator algebra 
 and in Appendix~\ref{App:B} we obtain the Feynman propagator for
the covariant Maxwell theory. In Appendix~\ref{app:FP},
we derive the Faddeev-Popov operator associated with
the $R_\xi$ gauge.
\section{The methodology}\label{sectionII}
In this section, we develop a constructive method that relates the multiplicity of classical solutions to the emergence of dipole ghost sectors in the quantum theory. In particular, we show how degenerate solutions give rise to generalized modes, double poles in the propagator, and the associated dipole ghost structure. 
\subsection{Maxwell theory in $2+1$ dimensions}
Consider the Maxwell Lagrangian with a gauge fixing term 
\begin{align}
\mathcal{L}=-\frac{1}{4}F_{\mu\nu}F^{\mu\nu}
-\frac{1}{2\xi} \big(\partial\cdot A\big)^2\,,
\label{maxwell3}
\end{align}
where $F^{\mu\nu}\equiv \partial^{\mu}A^{\nu}-\partial^{\nu}A^{\mu}$ 
is the electromagnetic field strength tensor, $\xi$ is the gauge 
fixing parameter, and we use the signature $\eta_{\mu\nu}=\text{diag.}(+,-,-)$. 

The field equations obtained from variation of the Lagrangian~\eqref{maxwell3} are given by
\begin{align}\label{eqmov}
\Bigg[ \eta^{\mu \nu} \Box  
-\left(1-\frac{1}{\xi}\right)\partial^\mu \partial^\nu 
    \Bigg]  A_\nu (x) =0\,.
\end{align}
Taking the derivative $\partial_\mu$ above, the Lorentz 
condition $\Lambda:=\partial_{\mu}  A^{\mu}$ is found to 
satisfy the massless 
wave equation 
\begin{align}
\Box \Lambda=0\,.
\end{align}

To solve the equations of motion, consider the following ansatz
\begin{align}\label{Ansatz_field}
    A_\nu(x)= \int \frac{\mathrm{d}^2k}{(2\pi)^2}u_\nu(t,\vec k) e^{i\vec k \cdot \vec  x}\,,
\end{align}
which, after substituted into Eq.~\eqref{eqmov}, yields the following two equations
\begin{subequations} \label{usual-fe}
\begin{align}
 & \vert\vec{k}\vert^2 u_0(t,\vec k) +\frac{1}{\xi}\ddot{u}_0 (t,\vec k) 
+ i\left(1-\frac{1}{\xi}\right)
k^j \notag \\&\hspace{6em} \times \dot{u}_j (t,\vec k) =0\,,  \\ &
 i\left(1-\frac{1}{\xi}\right)k^i   
\dot{u}_0(t,\vec k) +\eta^{i j} \ddot{u}_j(t,\vec k) 
+\Bigg[\eta^{i j} \vert\vec{k}\vert^2  
 \notag \\ &\hspace{6em}+\left(1-\frac{1}{\xi}\right)k^i k^j 
 \Bigg]  u_j (t,\vec k) =0 \,.
\end{align}
\end{subequations}
We decompose $u_j$ into vector and axial-vector components as 
\begin{eqnarray}\label{dir-decomposition-foton}
u_j(t,\vec k)=k_j f(t,\vec k)+\epsilon_{jl}k^l g(t,\vec k)\,, 
\end{eqnarray}
where $f(t,\vec k)$ and $g(t,\vec k)$ are 
functions to be determined. Substituting
into the equations of motion Eqs.~\eqref{usual-fe}, and taking into account that the vector 
and axial-vector sectors are linearly independent, we obtain the 
following system of equations
\begin{subequations}
\begin{align}
\vert\vec{k}\vert^2 u_0 +\frac{1}{\xi}\ddot{u}_0  - i\vert\vec{k}\vert^2\left(1-\frac{1}{\xi}\right)    
\dot{f}   &=0 \,,  \label{dipoleeq1} \\ 
i\left(1-\frac{1}{\xi}\right)\dot{u}_0 + 
\ddot{f}+ \frac{1}{\xi}  
\vert\vec{k}\vert^2 f&=0\,, \label{dipoleeq2} \\\ddot{g}  +\vert\vec{k}\vert^2  g &=0\,.
\end{align}
\end{subequations}
The last equation is decouple, and we 
can find the general solution for $g(t,\vec k)$, which is
\begin{align}
    g(t,\vec k)= c^{(1)}(\vec k) e^{-i\omega t}+c^{(2)}(\vec k) e^{i\omega t}\,,
\end{align}
where $\omega=|\vec{k}|$. For $u_0(t,\vec k)$ and $f(t,\vec k)$, we consider the solutions 
\begin{align}
    u_0(t,\vec k)&=A(\vec k) e^{-i \lambda t}\,, \\
    f(t,\vec k)&=B(\vec k) e^{-i \lambda t}\,,
\end{align}
where $\lambda$ is 
an eigenvalue to be determined. The system of 
equations \eqref{dipoleeq1} and \eqref{dipoleeq2} produce the dispersion equation
\begin{align}
\frac{1}{\xi}\big(\vert\vec{k}\vert^2-\lambda^2\big)^2 =0\,. \label{disp-usual}
\end{align}
Therefore, the solution $ \lambda=\pm \omega$  possess multiplicity two.
Since the system is second order in time derivatives, 
the corresponding general solution takes the form
\begin{subequations}\label{U_0F}
 \begin{align}
u_0(t,\vec k)=&\left(a^{(1)}(\vec k)+a^{(2)}(\vec k)t\right)e^{-i\omega t}\\
&\hspace{2em}+\left(a^{(3)}(\vec k)+a^{(4)}(\vec k)t\right)e^{i\omega t}\,, \notag \label{u0-foton} \\
f(t,\vec  k)=&\left(b^{(1)}(\vec k)+b^{(2)}(\vec k)t\right)e^{-i\omega t} \\
&\hspace{2em} +\left(b^{(3)}(\vec k)+b^{(4)}(\vec k)t\right)e^{i\omega t}\,.\notag  
\end{align}
\end{subequations}

For simplicity of the calculation, and given the reality condition 
imposed on the solution, we may restrict our analysis to the negative-frequency 
sector when solving the algebraic system. Substituting Eqs.~\eqref{U_0F} into the field equations~\eqref{dipoleeq1}  and~\eqref{dipoleeq2}, we obtain
\begin{align}
     a^{(1)}(\vec{k})&=\omega      
 b^{(1)}(\vec{k})-i \left(\frac{1+\xi}{1-\xi}\right) b^{(2)}(\vec{k}) \,, \label{a1b1}\\
    a^{(2)}(\vec{k})&= \omega b^{(2)}(\vec{k}) \label{a2b2}\,.
\end{align}
and therefore 
\begin{align}
  u_0(t,\vec k)=&\left(\omega      
 b^{(1)}(\vec{k})-i \left(\frac{1+\xi}{1-\xi}\right) b^{(2)}(\vec{k})\notag \right.   \\
    & \hspace{4em}\left. +\omega b^{(2)}(\vec{k}) t\right) e^{-i\omega t}  +\text{c.c.}\,, \label{u0-foton-3} \\
     f(t,\vec  k)&=\left(b^{(1)}(\vec{k})+b^{(2)}(\vec{k})t\right)e^{-i\omega t}
     +\text{c.c.}\,. \label{f-foton-3}    
\end{align}
Also, considering~\eqref{dir-decomposition-foton} and the previous result, we have
\begin{align}
u_j(t,\vec k)=&  k_j \Big(b^{(1)}(\vec{k}) +b^{(2)}(\vec{k}) 
t\Big)e^{-i\omega t} \notag\\
&+\epsilon_{jl}k^l c^{(1)}(\vec{k})
e^{-i\omega t}+\text{c.c.}\,.
\end{align}
Finally, from Eq.~\eqref{Ansatz_field} the classical gauge field $\tilde A_{\mu}(x)$ can be written
as the sum 
\begin{align}
   \tilde A_{\mu}(x)=\tilde A^{\text{G}}_\mu(x)+\tilde A^{\text{D}}_\mu(x)+\tilde
   A^{\text{T}}_\mu(x)\,,
\end{align}
where
\begin{subequations} \label{campos-clasicos}
\begin{align}
\tilde A^{\text{G}}_\mu(x)&=-i \int \frac{\mathrm{d}^2k}{(2\pi)^2} k_\mu
\Big( a(\vec k) e^{-i k\cdot x}\notag \\ &\hspace{8em} - a^*(\vec k) e^{ik\cdot x}\Big)_{k_0=\omega }\,, \\
\tilde A^{\text{D}}_\mu(x)&= \int \frac{\mathrm{d}^2k}{(2\pi)^2} 
\Big(v_\mu(t,\vec k)b(\vec k) e^{-i k\cdot x}+v_\mu^*(t,\vec k) \notag \\ 
&\hspace{8em}\times b^*(\vec k) e^{ik\cdot x}\Big)_{k_0=\omega }\,, \\
\tilde A^{\text{T}}_\mu(x)&= \int \frac{\mathrm{d}^2k}{(2\pi)^2} \epsilon_{\alpha
\mu \beta}n^\alpha k^\beta \Big( c(\vec k) e^{-i k\cdot x}\notag \\ &\hspace{8em}+  
c^*(\vec k) e^{ik\cdot x}\Big)_{k_0=\omega }\,,
\end{align}
\end{subequations}
with the time-dependent four-vector
\begin{align}
v_\mu(t,\vec k)&:= k_\mu  t-i\bigg(\frac{1+\xi}{1-\xi} \bigg) n_\mu\,,  \label{v-def}
\end{align}
and the purely timelike one
\begin{align}\label{ndefinition}
n_\mu&:=(1,0,0)\,.
\end{align}
Above we have made the identification $b^{(1)}(\vec k)\to a(\vec k)$ ,
$b^{(2)}(\vec k)\to b(\vec k)$, and
$c^{(1)}(\vec k)\to c(\vec k)$.

One can verify that the solutions $\tilde{A}_\mu^{\text{G}},~\tilde{A}_\mu^{\text{D}}$, and $\tilde{A}_\mu^{\text{T}}$ are, in fact, linearly independent solutions of the equation of motion, Eq.~\eqref{eqmov}. While it is straightforward to verify that the terms proportional to $k_\mu$ and $\epsilon_{\alpha\mu\beta}n^\alpha k^\beta$ satisfy the equation of motion, the solution proportional to $v_\nu(t,\vec{k})$ requires additional steps because of its functional dependence. To make this explicit, one may use the identities:
\begin{align}
    \Big[ \eta^{\mu \nu} \Box   
\Big]  v_\nu e^{-ik\cdot x}
&=  -2i (k\cdot n)   k^\mu    e^{-ik\cdot x} \,, \\
\Big[\partial^\mu \partial^\nu 
    \Big]  v_\nu e^{-ik\cdot x}&=\frac{2i\xi}{1-\xi}k^\mu(k\cdot n)   e^{-ik\cdot x} \,.
\end{align}
\subsection{Quantization}
The quantization 
can be performed directly by promoting the mode coefficients 
of the fields~\eqref{campos-clasicos} to creation and annihilation operators. The resulting 
field operators are given by
\begin{align}
    A_{\mu}(x)&=A^{\text{G}}_\mu(x)+A^{\text{D}}_\mu(x)+A^{\text{T}}_\mu(x)\,,
\end{align} 
with
\begin{subequations}\label{Fields_Max}
\begin{align}    \label{fieldcomponents}
A^{\text{G}}_\mu(x)&= -i\int \frac{\mathrm{d}^2k}{(2\pi)^2} k_\mu
\Big( a_{{k}} e^{-i k\cdot x} - a^\dagger_{{k}} e^{ik\cdot x}\Big)_{k_0=\omega } \,,\\
A^{\text{D}}_\mu(x)&= \int \frac{\mathrm{d}^2k}{(2\pi)^2} 
\Big(v_\mu(t,k)b_{{k}} e^{-i k\cdot x}+v_\mu^*(t,k)\notag  \\&\hspace{10em}\times
b^\dagger_{{k}} e^{ik\cdot x}\Big)_{k_0=\omega } \,,\\
A^{\text{T}}_\mu(x)&= \int \frac{\mathrm{d}^2k}{(2\pi)^2}
\epsilon_{\alpha \mu \beta}n^\alpha k^\beta \Big(c_{{k}} e^{-i k\cdot x}+ 
c^\dagger_{{k}} e^{ik\cdot x}\Big)_{k_0=\omega }\,.
\end{align}
\end{subequations}
By imposing the equal-time commutation relations
\begin{align}
\big[ A_\mu(x),\Pi^\nu(y)\big]&=i 
\delta^\nu_\mu \delta^{(2)}(\vec{x}-\vec{y})\,,
\end{align}
with all other commutators vanishing, 
and with 
\begin{align}
    \Pi^\mu&= - F^{0\mu}  -\frac{1}{\xi} 
    \big(\partial\cdot A\big)\eta^{0\mu} \,,
\end{align}
the algebra of 
creation and annihilation operators is found to be, 
\begin{subequations}\label{algebraoscilator}
\begin{align}
\big[a_{{k}},a_{{q}}^\dagger\big]&= (2\pi)^2 
\frac{(1+\xi)}{4\omega^3}  \delta^{(2)}(\vec{k}-\vec{q}), \\
\big[a_{{k}},b_{{q}}^\dagger\big]&=- (2\pi)^2
\frac{(1-\xi)}{4\omega^2} 
\delta^{(2)}(\vec{k}-\vec{q}), \\
\big[c_{{k}},c_{{q}}^\dagger\big]&= (2\pi)^2  \frac{1}{2 \omega^3  
} \delta^{(2)}(\vec{k}-\vec{q})\,.
\end{align}
\end{subequations}
A systematic derivation of the algebra is presented in Appendix~\ref{App:A}.

From the operator in momentum space 
\begin{align}
S_{\mu \nu}= -\eta_{\mu \nu} k^2  
+\left(1-\frac{1}{\xi}\right)k_\mu k_\nu \,,
\end{align}
the corresponding propagator,
satisfying $S_{\mu \nu} G^{\nu \alpha} =-i\delta_{\mu}^{\alpha}$, is given by
\begin{align}
G_{\mu \nu}= i\left(\frac{\eta_{\mu\nu} }{k^2}+\frac{  (\xi-1) }{(k^2)^2} k_\mu k_\nu  \right)\,, \label{Usual-Propagator}
\end{align}
which displays a double pole in the second term.
A derivation of the Feynman propagator 
from the definition in terms of temporal ordered field operators is presented in Appendix~\ref{App:B}.
\subsection{The Maxwell dipole ghost}
In this section, we explicitly identify the dipole ghost sector 
hidden in the quantized gauge field. Starting from 
the decomposition Eq.\eqref{Fields_Max},
we perform a field redefinition that separates 
the generalized mode associated with the degenerate sector. 
Our goal is to rewrite the gauge field in the form
\begin{align}
    A_{\mu}=A^{F}_{\mu}+\partial_\mu \chi\,,
\end{align}
where $A^{F}_{\mu}$ contains the ordinary propagating 
degrees of freedom in the Feynman gauge, while $\chi$ 
encodes the non diagonal sector of the theory. We will show 
that $\chi$ satisfies the fourth-order equation
\begin{align}\label{Dip_Ghost_Eq}
        \Box^2 \chi=0\,,
\end{align}
which is the defining equation of a massless dipole ghost 
field. Consequently, the generalized solutions associated 
with the repeated roots of the dispersion equation \eqref{disp-usual} are 
entirely captured by $\chi$, making the dipole ghost structure 
manifest at the level of the field expansion.

Let us start rewriting the original field expansion in terms of the 
standard polarization basis of the photon field, being 
\begin{subequations}
    \begin{align}
    \epsilon_\mu^{(0)}&=n_\mu \,, \\
    \epsilon_\mu^{(1)}&=\frac{1}{\omega}\epsilon_{\alpha\mu\beta} n^\alpha k^\beta \,,\\
    \epsilon_\mu^{(2)}&= \frac{1}{\omega}\left(k_\mu-\omega_k n_\mu\right)\,,
\end{align}
\end{subequations}
satisfying
\begin{align}
    \sum_{\lambda=0}^{2}\epsilon_\mu^{(\lambda)}\epsilon_\nu^{(\lambda')}\eta_{\lambda\lambda'}
    =\eta_{\mu\nu},
\end{align}
where $\eta_{\lambda\lambda'}=\mathrm{diag}(1,-1,-1)$ 
corresponds to the metric in polarization space and the timelike vector 
$n_{\mu}$ is given by~\eqref{ndefinition}. The corresponding 
creation and annihilation operators, $a_k^{(\lambda)}$ and $a_k^{\dagger(\lambda)}$, 
satisfy the canonical algebra
\begin{align}
    [a_{k}^{(\lambda)},{a}_q^{\dagger(\lambda')}]
    =-\eta^{\lambda\lambda'}\delta^{(2)}(\vec{k}-\vec{q}).
\end{align}
The polarization vectors can be combined in the form 
\begin{align}
    k_\mu =\omega\left(\epsilon_\mu^{(0)}+\epsilon_\mu^{(2)}\right),
\end{align}
and
\begin{align}
    \epsilon_{\alpha\mu\beta}n^\alpha k^\beta=\omega\epsilon_\mu^{(1)}.
\end{align}
Therefore, the gauge and transverse sectors can be 
immediately rewritten in terms of the standard photon polarizations. The nontrivial step is the decomposition of the degenerate mode $A_\mu^{\mathrm D}$, whose polarization vector
\begin{align}
v_\mu(t,k) = k_\mu t - i \left(\frac{1+\xi}{1-\xi}\right)n_\mu\,,
\end{align}
contains both a generalized eigenvector proportional 
to $k_\mu t$ and a temporal contribution proportional to $n_\mu$. 
As we shall show below, these two 
structures combine into a dipole field $\chi$ satisfying~\eqref{Dip_Ghost_Eq},
while the remaining degrees of freedom can be organized into 
ordinary photon modes with well-defined norm.

According to this, we redefine 
\begin{subequations}
\begin{align}
  a_{{k}} &=   iN_k\Big( (1+\alpha)  a_{{k}}^{(2)}- \alpha a_{{k}}^{(0)}\Big) \,,\\
  b_{{k}} &= \frac{i\alpha  N_k }{\omega} \big( a_{{k}}^{(2)}-a_{{k}}^{(0)}\big)  \,,\\
c_{k}&=(2\pi)^2\frac{N_k}{\omega} a^{(1)}_k\,,
\end{align}
\end{subequations}
with 
\begin{align}
    \alpha&=-\frac{(1-\xi)}{   4 } \,,\\
        N_{k}&= \frac{1}{\sqrt{2\omega(2\pi)^2}}.
\end{align}

After some straightforward calculations, we finally arrive at
\begin{align}
A_{\mu}=A_{\mu}^F+\partial_\mu \chi\,,
\end{align}
with 
\begin{align}
A_{\mu}^F=\int \mathrm{d}^2k N_{k}\sum_{\lambda=0}^2
\epsilon_{\mu}^{(\lambda)}\left(a_k^{(\lambda)} e^{ik\cdot x}
+a_k^{\dagger(\lambda)} e^{-ik\cdot x}\right)\,,
\end{align}
and
\begin{align}
 \chi(x)&=-i(1-\xi)\int  \mathrm{d}^2k  \frac{ N_k}{   2\omega} 
 \bigg[\bigg(  i \omega t+\frac{1}{2 }\bigg)    \lambda_k   e^{-i k\cdot x}\notag \\
 &\hspace{5em}+\bigg( i\omega    t- \frac{1}{2}\bigg)   
 \lambda_k^{\dagger}e^{ik\cdot x} \bigg]_{k_0=\omega }
\,,
\end{align}
where 
\begin{align}
   \lambda_k=a_k^{(2)}-a_k^{(0)}\,,
\end{align}
which is in agreement with Ref.~\cite{Greiner:1996zu}.

One can check that 
\begin{align}
\Box\left[
\left(i\omega t+\frac{1}{2}\right)e^{-ik\cdot x}
\right]&=
2\omega^2\,e^{-ik\cdot x}\,, \label{45}
\\
\Box\left[
\left(i\omega t-\frac{1}{2}\right)e^{ik\cdot x}
\right]
&=
-2\omega^2\,e^{ik\cdot x}\,, \label{46} 
\end{align}
and therefore~Eq.\eqref{Dip_Ghost_Eq} is satisfied. 
\section{The higher-order CS theory}\label{SectionIII}
We now apply the methodology developed in the previous sections to explicitly 
identify the dipole ghost sector of the extended CS theory. As we 
shall see, the gauge dependence of the degenerate sector differs significantly 
from that encountered in Maxwell theory. In the standard covariant Maxwell theory, 
as well as in the higher-order CS model with $\mu=0$~\cite{Avila:2019xdn}, 
the degenerate gauge contribution is eliminated in the Feynman gauge, $\xi=1$. 
By contrast, in the extended CS theory considered here, where both 
the conventional and higher-derivative CS terms are present, the 
degenerate modes disappear in the Landau gauge limit, $\xi\rightarrow0$~\cite{Avila:2025jxv}. 
This behavior reflects the nontrivial interplay between the two topological 
sector its higher derivative extension and the gauge-fixing procedure.
\subsection{Classical CS multiplicity}
The extended CS theory is described by the Lagrangian
\begin{align}\label{Lagrangian_parts}
\mathcal{L}_{\text{eCS}}&=-\frac{1}{4\gamma}F_{\mu\nu}
F^{\mu\nu}  -\frac{1}{2\xi}(\partial_\mu A^{\mu})^2	+ \frac{1}{2}\epsilon^{\alpha\beta\gamma} 
    A_\alpha \notag \\& \hspace{6em}\times \left(  \mu+g \Box  \right)
	\partial_\beta  A_\gamma \,,
\end{align}
where $\mu$ is the coefficient of the conventional CS 
term, while $g$ controls the higher-derivative extension of the model.

In the extended model the Lagrangian~\eqref{Lagrangian_parts}, 
the field equations of motion are
\begin{align}\label{G-F-E}
\Bigg[\eta^{\mu \nu} \frac{\Box }{\gamma}
-\left(\frac{1}{\gamma}-\frac{1}{\xi}\right)\partial^\mu \partial^\nu 
 +\epsilon^{\mu \beta\nu} \left( \mu+ g\Box \right)
\partial_\beta    \Bigg]  A_\nu (x) =0\,,
\end{align}
which we proceed to solve by using the following ansatz
\begin{eqnarray}
    A_{\nu} (t,\vec{x})= \int\frac{\mathrm{d}^2k}{(2\pi)^2}
    w_{\nu}(t,\vec{k}) e^{-i \vec{k}\cdot\vec{x}},
\end{eqnarray}
where $w_\nu(t,\vec{k})$ are arbitrary functions. They satisfy
\begin{subequations}
\begin{align}
 & \frac{1}{\xi}\ddot{w}_0+\frac{1}{\gamma}|\vec{k}|^2w_0 +ig  \epsilon^{ij}  
 k_i \ddot{w}_j-i\bigg(\frac{1}{\gamma}-\frac{1}{\xi}\bigg)
   k^i\dot{w}_i   \notag   \\ & \hspace{4em}+  i\epsilon^{ij}  
   \big(\mu+g |\vec{k}|^2 \big) k_i w_j  =0 \,, 
\end{align}
and
\begin{align}
 & ig\epsilon^{ij}k_j \ddot{w}_0-i\bigg(\frac{1}{\gamma}-\frac{1}{ \xi}\bigg)k^i 
 \dot{w}_0+ i\epsilon^{ij} (\mu+g |\vec{k}|^2)k_j w_0 \notag \\&\hspace{4em} 
 - g\epsilon^{ij} \dddot{w}_j+ \frac{1}{\gamma}\ddot{w}^i- \epsilon^{ij} 
 (\mu+g | \vec{k}|^2) \dot{w}_j \notag \\
 &\hspace{4em} +\frac{1}{ \gamma}  
 |\vec{k}|^2 w^i  +\bigg(\frac{1}{\gamma}-\frac{1}{ \xi}\bigg)k^i  k^j w_j   =0 \,.
\end{align}
\end{subequations}
From these equations, we can decompose in a contribution along $k^i$ 
and another contribution along $\epsilon^{ij}k_j$ with
\begin{align}
    w_j(t,\vec{k})=r(t,\vec{k})k_j+h(t,\vec{k})\epsilon_{jl}k^l
    \,. \label{uj-CS}
\end{align}
Replacing above, we obtain three equations 
\begin{subequations}  \label{Eq-CS}
\begin{align}
\frac{1}{\xi}\ddot{w}_0+\frac{1}{\gamma}\vert\vec{k}\vert^2
w_0+ig \vert\vec{k}\vert^2 \ddot{h} +i\bigg(\frac{1}{\gamma}-\frac{1}{\xi}\bigg)\vert\vec{k}\vert^2\dot{r}
& \notag  \\& \hspace{-14em}+  i    \big(\mu+g\vert\vec{k}\vert^2 
\big)\vert\vec{k}\vert^2 h  =0 \,,
\end{align}
\begin{align}
-i\bigg(\frac{1}{\gamma}-\frac{1}{ \xi}\bigg)  
\dot{w}_0 + g\dddot{h} + \frac{1}{\gamma}\ddot{r}+ 
(\mu+g \vert\vec{k}\vert^2)  \dot{h} & \notag \\   &
     \hspace{-12em} +\frac{1}{ \xi} \vert\vec{k}\vert^2 r    =0\,, 
\end{align}
and
\begin{align}
ig \ddot{w}_0+ i(\mu+g \vert\vec{k}\vert^2) w_0- 
g\dddot{r} + \frac{1}{\gamma}\ddot{h}& \notag \\ & \hspace{-14em}- 
(\mu+g \vert\vec{k}\vert^2) \dot{r}  +\frac{1}{ \gamma} 
\vert\vec{k}\vert^2    h     =0 \,.
\end{align}
\end{subequations}
Next, we consider the ansatz 
\begin{subequations}
\begin{align}
    w_0(t, \vec k)&=D(\vec k) e^{i\lambda t}\,, \\
    w_j(t,\vec{k})&=E(\vec k) e^{i\lambda t}k_j+F(\vec k)
    e^{i\lambda t}\epsilon_{jk}k^k \,,
\label{uj-CS}
\end{align}    
\end{subequations}
and replacing in~\eqref{Eq-CS},
we obtain a system of three equations
giving the dispersion equation
\begin{align}
 \Big(\big(\lambda^2&-\vert\vec{k}\vert^2\big)
 -\gamma^2 \big(\mu-g(\lambda^2-\vert\vec{k}\vert^2)\big)^2\Big) \\&\hspace{8em}  \times
\big(\lambda^2-\vert\vec{k}\vert^2\big)^2=0\notag \,.
\end{align}
Solving, we have the massive solutions  
\begin{align}
    \lambda_1&= \pm  \sqrt{\vert\vec{k}\vert^2+m_1^2} =\pm \omega_1\,,\\
    \lambda_2&= \pm \sqrt{\vert\vec{k}\vert^2+M_2^2}=\pm W_2\,,
\end{align}
with 
\begin{align} \label{bosonmasses1}
m_1= &\frac{\sqrt{1+4\gamma^2 \mu g}-1}{2 \gamma g} \,,  \\ \label{bosonmasses2}
M_2=& \frac{\sqrt{1+4 \gamma^2 \mu g}+1}{2\gamma g} \,.
\end{align}
We also have a photon solution with multiplicity two
\begin{align}
    \lambda_3= \pm \vert\vec{k}\vert=\pm\omega_0\,.
\end{align}
The number of independent coefficients can be reduced through a 
two-step procedure. First we impose reality conditions on the gauge field
and therefore we write 
\begin{align}
    w_0(t, \vec k)=&\big(d^{(1)}(\vec{k})+d^{(2)}(\vec{k}) t\big) e^{-i\omega_0 t} 
    +d^{(5)}(\vec{k}) e^{-i\omega_1 t} \notag \\
    &  +d^{(7)}(\vec{k}) e^{-iW_2 t}+\text{c.c} \,, \label{u0-CS}
    \end{align}
together with
\begin{align}
    r(t, \vec k)=&\big(e^{(1)}(\vec{k})+e^{(2)}(\vec{k}) t\big) 
    e^{-i\omega_0 t}+e^{(5)}(\vec{k}) e^{-i\omega_1 t}
    \notag \\
    &+e^{(7)}(\vec{k}) e^{-iW_2 t}+\text{c.c}\,, \label{r-CS}
    \\
    h(t, \vec k)=&\big(f^{(1)}(\vec{k})+f^{(2)}(\vec{k}) t\big) e^{-i\omega_0 t}
    +f^{(5)}(\vec{k}) e^{-i\omega_1 t}\notag\\
    &+f^{(7)}(\vec k) e^{-iW_2 t}+\text{c.c}\,. \label{h-CS}
\end{align}
Next, we solve the equations associated with the negative-frequency part
of the degenerate mode $\omega_0$, obtaining
\begin{subequations}\label{M_EQ}
\begin{eqnarray}
    d^{(1)}(\vec{k})&=&-\vert\vec{k}\vert e^{(1)}(\vec{k})-i e^{(2)}(\vec{k})\,, \label{A1B1B2}\\
    d^{(2)}(\vec{k})&=&-\vert\vec{k}\vert e^{(2)}(\vec{k})\,, \label{A2B2-2}\\
    f^{(1)}(\vec{k})&=&-\frac{2}{\mu\xi} e^{(2)}(\vec{k})\,, \label{C1B2} \\
    f^{(2)}(\vec{k})&=&0 \,.\label{C2}
\end{eqnarray}
\end{subequations}
Replacing Eqs.~\eqref{M_EQ} in 
Eq.~\eqref{u0-CS} and~\eqref{uj-CS}, we arrive at
\begin{align}
w_0(t,\vec{k})&=-\big(\vert\vec{k}\vert e^{(1)}(\vec{k})+(i  +\vert\vec{k}\vert t)e^{(2)}(\vec{k})  \big) e^{-i\omega_0 t}\notag \\ 
&+d^{(5)}(\vec{k}) e^{-i\omega_1 t} +d^{(7)}(\vec{k}) e^{-iW_2 t}+\text{c.c} \,,  
\end{align}
and 
\begin{align}
w_j(t,\vec{k})&=\Big[\big(e^{(1)}(\vec{k})+e^{(2)}(\vec{k}) t\big) e^{-i\omega_0 t}     +e^{(5)}(\vec{k}) e^{-i\omega_1 t} \notag \\
&+e^{(7)}(\vec{k}) e^{-iW_2 t}  \Big]k_j+\Big[  -\frac{2}{\mu\xi} e^{(2)}(\vec{k}) e^{-i\omega_0 t}
\notag \\
&+f^{(5)}(\vec{k}) e^{-i\omega_1 t} +f^{(7)}(\vec{k}) e^{-iW_2 t}\Big]\epsilon_{jk}k^k +\text{c.c}\,.
\end{align}

The general solution to the system corresponds to the 
sum of four fields with one degree of freedom each, given by
\begin{align}
  \tilde  A_\mu^{\text{CS}}= \tilde A_\mu^{D,\text{CS}}+\tilde
  A^{G,\text{CS}}_\mu+ \tilde {\bar{A}}_\mu+\tilde G_\mu  \,,
\end{align}
with
\begin{align}
 \tilde   A_\mu^{G,\text{CS}}(x)&=-i\int \frac{\mathrm{d}^2k}{(2\pi)^2} 
 \frac{1}{2\omega_0 }k_\mu \Big[  {e}(\vec k)e^{-ik\cdot x}\notag \\ &\hspace{6em} -e^*(\vec k)e^{ik\cdot x}
 \Big]_{k_0=\omega_0} \,, \\
    \tilde A^{D,\text{CS}}_\mu(x)&=\int\frac{\mathrm{d}^2k}{(2\pi)^2} \frac{1}{2\omega_0  }\bigg[ v_\mu(t,k) d(\vec k)e^{-ik\cdot x}\notag \\& \hspace{5.5em} +v_\mu^*(t,k)d(\vec k)^* e^{ik\cdot x} 
    \bigg]_{k_0=\omega_0} \,,\\
    \tilde{\bar{A}}_\mu(x)&=\int \frac{\mathrm{d}^2
    \vec{k}}{(2\pi)^2}\frac{1}
    {2\omega_1\beta}\bigg[\varepsilon_\mu^{(-)}
    c^{(-)}(\vec k)e^{-i k\cdot x}+ \notag \\& \hspace{6em} \varepsilon_\mu^{(-)*}
    c^{(-)*}(\vec k)e^{i k\cdot x}   \bigg]_{k_0=\omega_1} \,, \\
  \tilde G_\mu(x)&=\int \frac{\mathrm{d}^2\vec{k}}{(2\pi)^2}
    \frac{1}{2W_2\beta}\bigg[\varepsilon_\mu^{(+)}
    c^{(+)}(\vec k)e^{-i k\cdot x}+ \notag \\& \hspace{6em} \varepsilon_\mu^{(+)*}
    c^{(+)*}(\vec k)e^{i k\cdot x}   \bigg]_{k_0=W_2} \,,
\end{align}
where 
\begin{align}
    v_\mu&=\mu t k_\mu+i\mu n_\mu -\frac{2}{\xi} \epsilon_{\mu \rho\sigma}k^\rho n^\sigma \,, 
\end{align}  
and $n$ given in Eq.~\eqref{ndefinition}, and $\beta:=\frac{\sqrt{1+4\gamma^2\mu g}}{2\gamma}$. 
Above me have made the identification $e^{(1)}\to e$ , $e^{(2)}\to d$, $f^{(5)}\to c^{(-)}$
and $f^{(7)}\to c^{(+)}$ .

The field $\tilde A_\mu^{D,CS}$ corresponds to a photon solution 
with multiplicity two in the eigenvalue problem, $\tilde {\bar{A}}_\mu$ corresponds to a massive photon solution, $\tilde G_\mu$ corresponds to a massive ghost solution, and 
$\tilde A_\mu^{G,CS}$ to a longitudinal mode.
\subsection{CS quantization}
The canonical quantization of the Chern-Simons model was carried out in Ref.~\cite{Avila:2025jxv}, where the 
complete constraint structure was determined. In particular, all constraints were shown to be second class, 
leading naturally to the introduction of Dirac brackets. Furthermore, it was demonstrated that the quantized 
fields satisfy equal-time commutation relations consistent with the corresponding Dirac bracket algebra, providing 
a nontrivial consistency check of the quantization procedure.

Following this construction, the field expansion takes the form

\begin{align}
    A_\mu^{\text{CS}}=  A_\mu^{D,\text{CS}}+
  A^{G,\text{CS}}_\mu+  {\bar{A}}_\mu+ G_\mu  \,,
\end{align}
with
\begin{align}
    A^{D,\text{CS}}_\mu(x)&=\int\frac{\mathrm{d}^2k}{(2\pi)^2} \frac{1}{2\vert\vec{k}\vert  }
    \bigg[ v_\mu(t,k) \bar a_{{k}}e^{-ik\cdot x} \notag \\& \hspace{6em}+v_\mu^*(t,k)\bar a_{{k}}^\dagger e^{ik\cdot x}
    \bigg]_{k_0=\omega_0} \,, \\
   A_\mu^{G,\text{CS}}(x)&=-i\int \frac{\mathrm{d}^2k}{(2\pi)^2}  \frac{k_\mu }{2\vert\vec{k}\vert}
   \Big[ \bar b_{{k}}e^{-ik\cdot x}-\bar b_{{k}}^\dagger e^{ik\cdot x}\Big]_{k_0=\omega_0} \,, \\
    \bar{A}_\mu(x)&=\int \frac{\mathrm{d}^2
    \vec{k}}{(2\pi)^2}\frac{1}
    {2\omega_1\beta}\bigg[\varepsilon_\mu^{(-)}
    c_{{k}}^{(-)}e^{-i k\cdot x}\notag \\& \hspace{6em} + \varepsilon_\mu^{(-)*}
    c_{{k}}^{(-)\dagger}e^{i k\cdot x}   \bigg]_{k_0=\omega_1} \,, \\
  G_\mu(x)&=\int \frac{\mathrm{d}^2\vec{k}}{(2\pi)^2}
    \frac{1}{2W_2\beta}\bigg[\varepsilon_\mu^{(+)}
    c_{{k}}^{(+)}e^{-i k\cdot x}\notag \\& \hspace{6em}+ \varepsilon_\mu^{(+)*}
    c_{{k}}^{(+)\dagger}e^{i k\cdot x}   \bigg]_{k_0=W_2} \,.
\end{align}
The algebra of the creation and annihilation operators is given by
\begin{subequations} \label{CS-algebra}
\begin{align}
   \big[ {\bar b}_{k},\bar a^{\dagger}_{q}\big]&=(2\pi)^2  \frac{\xi }{\mu}  \delta^{(2)}(\vec{k}-\vec{q})] \,, \label{78a}\\
    \big[ {\bar b}_{k},\bar b_{q}^\dagger\big]&= (2\pi)^2 \bigg[-\frac{ 2\vert\vec{k}\vert  }{\gamma \mu^2}+
    \frac{\xi}{ \vert\vec{k}\vert } \bigg] \delta^{(2)}(\vec{k}-\vec{q}) \,, \label{78b} \\
\big[c^{(+)}_{k},c^{(+)\dagger}_{q}\big]&=-(2\pi)^2 2 W_2\beta\delta^{(2)}(\vec{k}-\vec{q})\,, \\
\big[c^{(-)}_{k},c^{(-)\dagger}_{q}\big]&=(2\pi)^2 2\omega_1
    \beta\delta^{(2)}(\vec{k}-\vec{q}) \,,\label{39}
\end{align}
\end{subequations}
which has been derived in~\cite{Avila:2025jxv}.
\subsection{The CS dipole ghost}
In order to find the ghost dipole we 
introduce the change of variables
\begin{align} \label{new_op}
 \alpha_{{k}}  &=    \bar b_{{k}}- \frac{\mu}{2\vert\vec{k}\vert}\bar a_{{k}}\,, \\
 \beta_{{k}} &= \bar b_{{k}}+\bigg(\frac{2\vert\vec{k}\vert}{\xi\gamma\mu}- \frac{\mu}{2\vert\vec{k}\vert} \bigg)\bar a_{{k}} \label{new_op2}  \,,
\end{align}
which diagonalizes the algebra~\eqref{78a} and \eqref{78b}, giving
\begin{align}
    \big[ \alpha_{{k}},\alpha_{{q}}^\dagger\big]&= -(2\pi)^2 \frac{2\vert\vec{k}\vert}{\gamma\mu^2}\delta^{(2)}(\vec{k}-\vec{q})\,, \\
     \big[ \beta_{{k}},\beta_{{q}}^\dagger\big]&= (2\pi)^2 \frac{2\vert\vec{k}\vert}{\gamma\mu^2}\delta^{(2)}(\vec{k}-\vec{q}) \,,\\
      \big[ \alpha_{{k}},\beta_{{q}}^\dagger\big]&= \big[ \beta_{{k}},\alpha_{{q}}^\dagger\big]=0\,.
\end{align}
In terms of the new operators~\eqref{new_op} and~\eqref{new_op2}, we have
\begin{align}
    & A_{\mu}^{D,\text{CS}}(x)+A_\mu^{G,\text{CS}}(x)   =\int\frac{\mathrm{d}^2k}{(2\pi)^2} \frac{\gamma\mu }{2\vert\vec{k}\vert}\frac{\epsilon_{\mu \rho\sigma}k^\rho n^\sigma}
     { \vert\vec{k}\vert  } \notag  \\& \times   \Big[   \big( \alpha_{{k}}-\beta_{{k}}\big) e^{-ik\cdot x}  
     +  \big( \alpha^\dagger_{{k}}-\beta^\dagger_{{k}}\big) e^{ik\cdot x}  \Big]_{k_0=\omega_0} \notag \\
   &  
      +\int \frac{\mathrm{d}^2k}{(2\pi)^2}  \frac{1}{2\vert\vec{k}\vert} 
      \Big[  (-ik_\mu) \alpha_{{k}}  e^{-ik\cdot x}+ (ik_\mu)\alpha^\dagger_{{k}}e^{ik\cdot x}\Big]_{k_0=\omega_0}\notag \\
   &
  + \partial_\mu\chi_{CS} \,,
\end{align}
where
\begin{align}
  &  \chi_{CS}(x):=-\xi\int\frac{\mathrm{d}^2k}{(2\pi)^2} \frac{\gamma\mu^2}{4\vert\vec{k}\vert^2  } \bigg[   \bigg(it+   \frac{1}{ 2\vert\vec{k}\vert }\bigg)   \\&  \times \big( \alpha_{{k}}-\beta_{{k}}\big) e^{-ik\cdot x}+ \bigg(-i   t +   \frac{1}{ 2\vert\vec{k}\vert }  \bigg)  \big( \alpha^\dagger_{{k}}-\beta^\dagger_{{k}}\big) e^{ik\cdot x} \bigg]_{k_0=\omega_0}\notag \,,
\end{align}
is the dipole ghost.
One can check that $\Box^2 \chi_{CS}=0$ as anticipated 
using Eqs.~\eqref{45} and~\eqref{46}.

As mentioned above, the appearance of the solution $A_\mu^{\mathrm D}$ originates from the degeneracy of the massless modes satisfying $k^2=0$ in the longitudinal sector associated with the projector $\varepsilon^{(0)}\otimes\varepsilon^{(0)}$~\cite{Avila:2025jxv}. Such a degeneracy implies that the corresponding equations of motion are not completely described by ordinary eigenmodes. Instead, the solution space must be enlarged to include generalized modes, revealing an underlying non-diagonalizable structure known as a Jordan chain.

In the present case, the Jordan chain consists of two elements: the ordinary gauge mode $A_\mu^{\mathrm G}$ and the generalized mode $A_\mu^{\mathrm D}$. In the space of solutions they are related by successive applications of the wave operator,
\begin{align}
A^{\mathrm D}_{\mu}(x)
\overset{\Box}{\longmapsto}
A^{\mathrm G}_{\mu}(x)
\overset{\Box}{\longmapsto}
0.
\end{align}
The field $A_\mu^{\mathrm G}$ is therefore an ordinary eigenmode of the wave operator, while $A_\mu^{\mathrm D}$ plays the role 
of a generalized eigenmode. The length of the Jordan chain coincides with the multiplicity of the corresponding 
root of the characteristic polynomial, explaining the appearance of solutions proportional to $t\,e^{-i\omega_k t}$. 
In the Landau gauge the generalized mode disappears, reducing the chain to a single 
element and thereby eliminating the dipole sector.

Before concluding this section, it is worth commenting on the implementation 
of the Gupta-Bleuler prescription. An important observation is that, although 
the higher-order CS term modifies the spectrum of the 
theory and introduces a nontrivial degenerate sector, it does not alter 
the equation satisfied by the divergence of the gauge field. Indeed, the 
CS contribution 
is transverse and therefore drops out after taking the divergence of 
the equations of motion. As a consequence, the Lorentz condition
\begin{align}
\Lambda_{CS}(x):= \partial_\mu A^{\mu,\text{CS}}(x) \,,
\end{align}
continues to satisfy the free wave equation
\begin{align}
\Box \Lambda_{CS}(x)=0,
\end{align}
exactly as in ordinary covariant Maxwell theory.

The physical Hilbert space may therefore be defined through 
the standard Gupta-Bleuler condition
\begin{align}
\Lambda_{CS}^{(+)}(x)\,|\mathrm{phys}\rangle=0.
\end{align}
It is important to emphasize that the condition itself remains unchanged with respect to Maxwell theory. What changes is the operator content of the positive-frequency part of $\Lambda_{CS}$. In the Maxwell case, the Gupta-Bleuler condition removes the unphysical combination associated with the longitudinal and timelike sectors. In the present higher-order CS theory, however, $\Lambda^{(+)}_{CS}$ is expressed in terms of the operators $\alpha_{ k}$ and $\beta_{ k}$ that characterize the degenerate sector. Consequently, the Gupta-Bleuler prescription translates into a constraint on a particular combination of these modes, removing the corresponding unphysical excitations from the physical Hilbert space. Thus, while the emergence of 
the dipole ghost sector modifies the realization of the condition at the 
level of the oscillator algebra, the extended Gupta-Bleuler prescription itself 
remains valid and provides a consistent definition of the physical subspace.

\section{Spontaneous symmetry breaking in higher-order CS theory}~\label{SSBIV}
Having identified the dipole ghost sector as a consequence of the
degeneracy of the gauge spectrum, we now investigate how this structure
is modified by spontaneous symmetry breaking. To this end, we couple the
extended CS theory to a charged complex scalar field and
analyze the resulting vacuum structure. 

The gauge-invariant Lagrangian is
\begin{align}\label{modelssimmbreak}
\mathcal{L}=&\mathcal{L}_{\rm eCS}+(D_\mu\Phi)^*(D^\mu\Phi)-V(\Phi)\,,
\end{align}
where $\Phi$ is a charged scalar field,
\begin{equation}
D_\mu\Phi=(\partial_\mu+i e A_\mu)\Phi\,,
\end{equation}
denotes the covariant derivative, and the scalar self-interaction potential has the form
\begin{equation}
V(\Phi)=-\mu_H^2\,\Phi^*\Phi+\lambda(\Phi^*\Phi)^2 \,,
\end{equation}
with
\begin{equation}
\mu^2_H>0,\qquad\lambda>0\,.
\end{equation}

The scalar field develops a nonvanishing vacuum expectation value,
\begin{align}\label{vevcl}
v_0=\frac{\mu_H}{\sqrt{\lambda}}\,,
\end{align}
so, expanding the scalar field about this value, we have
\begin{align}
\Phi(x)=\frac{1}{\sqrt2}\left[v_0+h(x)+i\chi_G(x)
\right]\,,\label{PhiExpansion}
\end{align}
where $h$ and $\chi_G$ denote the Higgs and Goldstone fields, respectively.

Expanding the covariant kinetic term around the background field,
one finds a term proportional to
\begin{align}
(D_\mu\Phi)^\dagger(D^\mu\Phi) \supset -ev_0\,\chi_G\,\partial_\mu A^\mu \,.
\label{MixingTerm}
\end{align}
The appearance of this mixing indicates that the Lorenz gauge employed
in the symmetric phase no longer diagonalizes the quadratic
Lagrangian. A different gauge-fixing condition is therefore required to
obtain a diagonal propagator for the gauge and scalar fluctuations.

To eliminate this mixing, we adopt the usual $R_\xi$ gauge~\cite{Peskin:1995ev},
\begin{align}
\mathcal{L}_{\rm gf}^{R_\xi}=-\frac{1}{2\xi}\left(\partial_\mu A^\mu-\xi e v_0\,\chi_G\right)^2\,,
\end{align}
which exactly cancels the mixing contribution in
Eq.~(\ref{MixingTerm}). As a consequence, the quadratic Lagrangian becomes
diagonal in the gauge and scalar fields.

The adoption of the $R_\xi$ gauge also introduces a nontrivial
Faddeev-Popov sector. Since both the Maxwell and the extended
CS terms are gauge invariant, the corresponding
Faddeev-Popov operator is determined entirely by the gauge-fixing
functional and therefore coincides with that of the conventional
Abelian Higgs model~\cite{Peskin:1995ev}. Its derivation is presented in
Appendix~\ref{app:FP}.

The corresponding ghost Lagrangian is
\begin{align}
\mathcal{L}_{\rm FP}=\bar c\left(\Box+m_c^2(v_0)\right)c\,,\label{GhostLagrangian}
\end{align}
where
\begin{align}
m_c(v_0)=\xi^{1/2}e v_0   \,.
\end{align}

The quadratic gauge-field Lagrangian is therefore
\begin{align}\label{procalagrangian}
\mathcal{L}^{(2)}_A
=&-\frac{1}{4\gamma}F_{\mu\nu}F^{\mu\nu}
+\frac{1}{2}\epsilon^{\alpha\beta\gamma}
A_\alpha\left(\mu+g\Box\right)\partial_\beta A_\gamma
\nonumber\\
&+\frac{1}{2}m_\gamma^2A_\mu A^\mu
-\frac{1}{2\xi}(\partial_\mu A^\mu)^2\,,
\end{align}
where 
\begin{align}
    m_\gamma=ev_0\,,
\end{align} 
is the gauge-boson mass generated by the Higgs mechanism. It is important to emphasize 
that Eq.~\eqref{procalagrangian} differs from the corresponding quadratic Lagrangian in the 
symmetric phase only through the Higgs-generated Proca mass term. Therefore, spontaneous symmetry 
breaking leaves the tensorial structure of the gauge operator unchanged, modifying only its mass spectrum. 
As will be shown below, this seemingly modest modification is sufficient to lift the massless 
degeneracy responsible for the dipole ghost sector.

The scalar sector is given by
\begin{align}
\mathcal{L}^{(2)}_{\rm scalar}= & \frac12(\partial_\mu h)^2
-\frac12m_H^2(v_0)h^2
\nonumber\\&+ \frac12(\partial_\mu\chi_G)^2- \frac12m_{\chi_G}^2(v_0)\chi_G^2\,,
\end{align}
with
\begin{align}
m_H^2(v_0)&= -\mu_H^2+3\lambda v_0^2\,,\\m_{\chi_G}^2(v_0)
&= -\mu_H^2+\lambda v_0^2+\xi e^2v_0^2\,.
\end{align}
Notice that at the classical minimum $v_0$, given in \eqref{vevcl}, these reduce to
\begin{align}
m_H^2(v_0)=2\mu_H^2,\qquad m_{\chi_G}^2(v_0)=\xi m_\gamma^2\,.
\end{align}

The inverse propagator of the gauge field obtained from
Eq.~(\ref{procalagrangian}) is
\begin{align}
D^{-1}_{\mu\nu}(k)=&-\frac{\eta_{\mu\nu}}{\gamma}\left( k^2-\gamma m_\gamma^2(v_0) \right)
\notag\\& +\left( \frac1\gamma-\frac1\xi \right)
k_\mu k_\nu-i\mathcal M(k)\epsilon_{\mu\beta\nu}k^\beta\,,\label{procaInverseProp}
\end{align}
where
\begin{align}
\mathcal M(k)=\mu-gk^2\,. \label{m-de-cs}
\end{align}
Compared with the symmetric phase, the only modification of the quadratic
gauge operator is the Higgs-generated Proca mass. Nevertheless, this
additional mass scale completely changes the analytic structure of the
propagator. The repeated massless root responsible for the dipole ghost
sector is lifted into a set of ordinary massive poles, showing that
spontaneous symmetry breaking removes the degeneracy underlying the dipole
ghost sector.

The lifting of the gauge-sector degeneracy has important consequences
for the quantum vacuum of the theory. To quantify these effects, we now
compute the one-loop effective potential in the spontaneously broken
phase.
\subsection{One-loop effective potential}
With the quadratic spectrum of the spontaneously broken phase
established, we now turn to the analysis of the corresponding quantum
vacuum. We investigate the vacuum structure of the theory defined in
Eq.~\eqref{modelssimmbreak}, focusing on the role of the dipole ghost
sector through the one-loop effective potential. To this end, the
classical vacuum expectation value $v_0$ is promoted to an arbitrary
constant background field $v$~\cite{Jackiw:1974cv}, so that all field-dependent masses
appearing below are understood as functions of $v$. The physical
vacuum is then recovered by imposing the renormalization conditions at
$v=v_0$. At tree level, the scalar potential evaluated on the
background field is
\begin{align}
V^{(0)}(v)=-\frac12\mu_H^2 v^2+\frac14\lambda v^4\,.
\end{align}

The one-loop contribution is naturally decomposed into scalar, gauge,
and ghost sectors,
\begin{align}
V^{(1)}(v)= V^{(1)}_{\rm scalar}(v) + V^{(1)}_{\rm gauge}(v)
+ V^{(1)}_{\rm ghost}(v)\,.
\end{align}

The scalar contribution is
\begin{align}
V^{(1)}_{\rm scalar}(v)&=
\frac12\int\frac{\mathrm{d}^3k_E}{(2\pi)^3}\ln\left(
k_E^2+m_H^2(v)
\right)\notag \\ &+\frac12\int
\frac{\mathrm{d}^3k_E}{(2\pi)^3}\ln\left(
k_E^2+m_{\chi_G}^2(v)\right)\,,
\end{align}
with 
\begin{align}
m_H^2(v)&=-\mu_H^2+3\lambda v^2\,,\\
m_{\chi_G}^2(v)
&=
-\mu_H^2+\lambda v^2+\xi e^2v^2\,.
\end{align}

The gauge contribution is obtained from the quadratic operator \eqref{procaInverseProp},
\begin{align}
V^{(1)}_{\rm gauge}(v)=
\frac{i}{2}
\int\frac{\mathrm{d}^3k}{(2\pi)^3}
\ln\det \mathcal{D}^{{-1}}_{\mu\nu}(k)\,.
\end{align}

Now, taking into account that the determinant factorizes as
\begin{align}
\det \mathcal{D}^{{-1}}_{\mu\nu}(k)
=&
\left(m_\gamma^2(v)-\frac{k^2}{\xi}\right)\\
&\times
\left[
\frac{(k^2-\gamma m_\gamma^2(v))^2}{\gamma^2}
-k^2(\mu-gk^2)^2
\right]\,,\notag
\end{align}
and after performing the Wick rotation $k^2\rightarrow -k_E^2$, the gauge contribution becomes
\begin{align}\label{effectivepot}
V^{(1)}_{\rm gauge}(v)
=&\frac{1}{2}
\int\frac{\mathrm{d}^3k_E}{(2\pi)^3}
\ln\left(k^2_E+\xi m_\gamma^2(v)\right)
\nonumber \\
&+
\frac{1}{2}\sum_{j=1}^3
\int\frac{\mathrm{d}^3k_E}{(2\pi)^3}
\ln\left(k^2_E+M_j^2(v)\right)\,,
\end{align}
where the first logarithmic contribution originates from the gauge-dependent longitudinal mode, while 
the remaining three correspond to the massive poles of the gauge propagator, with 
the masses $M_j^2(v)$ given by
\begin{equation}
M^2_j(v)=
-\frac{1}{3g^2}
\left[
2\mu g+\frac{1}{\gamma^2}
+e^{\frac{2\pi i}{3}j} C
+e^{-\frac{2\pi i}{3}j} \frac{\Delta_0}{C}
\right],
\end{equation}
where $j=0,1,2,$
\begin{equation}
C=\sqrt[3]{\frac{\Delta_1+\sqrt{\Delta_1^2-4\Delta_0^3}}{2}}\,,
\end{equation}
and 
\begin{align}
\Delta_0&=\left(2\mu g+\frac{1}{\gamma^2}\right)^2-3g^2\left(\mu^2+\frac{2m_\gamma^2(v)}{\gamma}\right),
\\[1ex]\Delta_1
&=2\left(2\mu g+\frac{1}{\gamma^2}\right)^3\\&-9g^2
\left(2\mu g+\frac{1}{\gamma^2}\right)\left(\mu^2+\frac{2m_\gamma^2(v)}{\gamma}
\right)+27g^4m_\gamma^4(v).\notag
\end{align}

The ghost contribution follows directly from the quadratic ghost
Lagrangian,
Eq.~(\ref{GhostLagrangian}),
and is given by
\begin{align}
V^{(1)}_{\rm ghost}(v)=-\int\frac{\mathrm{d}^3k_E}{(2\pi)^3}\ln\left(k_E^2+\xi m_\gamma^2(v)\right).
\end{align}

Each of the previous contributions can be expressed in terms of the
elementary one-loop integral
\begin{align}
I(m)&\equiv\frac{1}{2}
\int\frac{\mathrm{d}^3k_E}{(2\pi)^3}
\ln\left(k^2_E+m^2\right),
\end{align}
it is sufficient to evaluate this quantity once. Using dimensional regularization, one finds~\cite{Branchina:2022jqc,Lee:1974fj}
\begin{align}
I(m)&=-\frac{\mu^{3-d}_{\rm DR}}{2(4\pi)^{d/2}}
\Gamma\!\left(-\frac{d}{2}\right)
\left[m^2\right]^{d/2},\\
&\stackrel{d\rightarrow3}{=}-\frac{1}{12\pi}m^3\,,
\end{align}
where $m$ denotes a generic mass associated with either a scalar or a 
gauge-field degree of freedom. The arbitrary scale $\mu_{\rm DR}$ 
is introduced to preserve the correct energy dimensions in $d$ dimensions 
and drops out after taking the physical limit $d\to3$. 

The one-loop effective potential obtained above determines the quantum
corrections to the vacuum structure of the theory. To preserve the
physical interpretation of the classical vacuum, we introduce a
renormalized effective potential by adding local counterterms whose
finite parts are fixed through suitable renormalization conditions at
the physical minimum, $v=v_0$. This procedure is described in the
following subsection.
\subsection{Renormalization of the effective potential}
To renormalize the one-loop effective potential, we introduce the
counterterm potential
\begin{align}\label{veffrenorma}
V_{\rm ct}(v)=\delta\Lambda-\frac12\delta\mu_H^2 v^2+\frac14\delta\lambda\,v^4 \,,
\end{align}
so, the renormalized effective potential takes the form
\begin{align}
V_{\rm eff}^{\rm ren}(v)=V^{(0)}(v)+V^{(1)}(v)+V_{\rm ct}(v)\,,
\end{align}
where
\begin{align}
V^{(0)}(v)=-\frac12\mu_H^2v^2+\frac14\lambda v^4\,,
\end{align}
and
\begin{align}
&V^{(1)}(v)= \\&\hspace{1em}-\frac1{12\pi}\Bigg[m_H^3(v)+m_{\chi_G}^3(v) -
\xi^{3/2}m_\gamma^3(v)+ \sum_{j=1}^{3}M_j^3(v)\Bigg]\,.\notag
\end{align}

We fix the counterterms by imposing the renormalization conditions
\begin{subequations}\label{renocond}
\begin{align}
V_{\rm eff}^{\rm ren}(v_0)&=V^{(0)}(v_0),\\\left.
\frac{dV_{\rm eff}^{\rm ren}}{dv}\right|_{v=v_0}&=0\,,\\
\left.\frac{d^2V_{\rm eff}^{\rm ren}}{dv^2}
\right|_{v=v_0}&=\left.\frac{d^2V^{(0)}}{dv^2}\right|_{v=v_0}\,.
\end{align}
\end{subequations}
These conditions ensure that the renormalized effective potential reproduces the 
classical vacuum energy, preserves the position of the symmetry-breaking minimum, and maintains the tree-level Higgs mass~\cite{Jackiw:1974cv}.

Solving these conditions gives
\begin{align}
\delta\lambda=\frac{1}{2v_0^2}\left[\frac{1}{v_0}\left.
\frac{\mathrm{d}V^{(1)}}{\mathrm{d}v}
\right|_{v=v_0}-\left.
\frac{\mathrm{d}^2V^{(1)}}{\mathrm{d}v^2}\right|_{v=v_0}\right]\,,
\end{align}
\begin{align}
\delta\mu_H^2=\frac12\left[
\frac{3}{v_0}\left.
\frac{\mathrm{d}V^{(1)}}{\mathrm{d}v}\right|_{v=v_0}-\left.\frac{\mathrm{d}^2V^{(1)}}{\mathrm{d}v^2}\right|_{v=v_0}\right]\,,
\end{align}
and
\begin{align}
\delta\Lambda=-V^{(1)}(v_0)+
\frac{5v_0}{8}\left.\frac{\mathrm{d}V^{(1)}}{\mathrm{d}v}\right|_{v=v_0}-\frac{v_0^2}{8}
\left.\frac{\mathrm{d}^2V^{(1)}}{\mathrm{d}v^2}\right|_{v=v_0}\,.
\end{align}

The renormalized effective potential is now completely determined.
Before proceeding to the numerical analysis, however, it is important to note
that, for field values in the vicinity of the symmetric configuration, the field-dependent Higgs and 
Goldstone squared masses become negative. Consequently, the one-loop effective potential develops an 
imaginary part, reflecting the instability of these field configurations rather than the energy of a 
stable vacuum. Following the
standard treatment of the effective potential, we therefore restrict our analysis to its real part and introduce
\begin{align}
\Delta V_{\rm eff}^{\rm ren}(v)
\equiv
V_{\rm eff}^{\rm ren}(v)-V_{\rm eff}^{\rm ren}(0)\,.
\end{align}

With this definition, the symmetric configuration is assigned zero
vacuum energy, and the stability of the broken phase is determined
directly by the sign and depth of the minimum of
$\Delta V_{\rm eff}^{\rm ren}(v)$.

\begin{figure}[h!]
	\includegraphics[width=8.7cm]{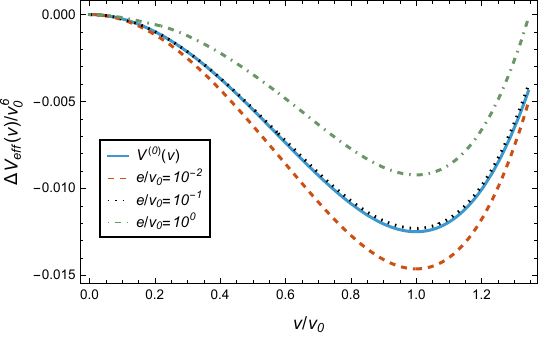}
	\caption{Renormalized effective potential shifted by its value at the symmetric
configuration, $\Delta V_{\rm eff}^{\rm ren}(v)=V_{\rm eff}^{\rm ren}(v)-V_{\rm eff}^{\rm ren}(0)$, 
normalized to $v_0^6$, as a function of $v/v_0$. The parameters are
fixed to $\gamma=1$, $\xi=1$, $\lambda/v_0^2=0.05$, $gv_0^2=1/50$, and
$\mu/v_0^2=7\times10^{-1}$. The solid curve denotes the tree-level
potential, while the dashed, dotted, and dot-dashed curves correspond
to $e/v_0=10^{-2},10^{-1},1$, respectively.}
	\label{Fig1}
\end{figure}

For the entire numerical analysis, we set
$\gamma=1$ and $\xi=1$.
In Fig.~\ref{Fig1}, we fix
 $\lambda/v_0^2=0.05$,
$gv_0^2=1/50$, and
$\mu/v_0^2=7\times10^{-1}$, and show
$\Delta V_{\rm eff}^{\rm ren}(v)/v_0^6$ as a function of
$v/v_0$ for three values of the normalized gauge coupling $e/v_0$.

To illustrate the effect of quantum corrections on the vacuum
structure, Fig.~\ref{Fig1} shows the shifted renormalized effective
potential, $\Delta V_{\rm eff}^{\rm ren}(v)$, normalized to $v_0^6$, as
a function of the dimensionless field $v/v_0$. By construction, the
potential vanishes at the symmetric configuration, $v=0$, allowing a
direct comparison between the energies of the symmetric and broken
phases. The solid black curve corresponds to the tree-level potential,
while the dashed blue, dotted red, and dot-dashed green curves include
the one-loop corrections for $e/v_0=10^{-2}$, $10^{-1}$, and
$1$, respectively. The remaining parameters are fixed at
 $\lambda/v_0^2=0.05$,
$gv_0^2=1/50$, and $\mu/v_0^2=7\times10^{-1}$.
The one-loop corrections preserve the position of the minimum at
$v=v_0$, as required by the renormalization conditions, while modifying
its depth. For weak gauge coupling, the quantum corrections deepen the
broken minimum relative to the tree-level prediction. As the gauge
coupling increases, gauge-field fluctuations progressively reduce the
depth of the minimum, although the broken phase remains energetically
favored throughout the range of couplings considered.

\begin{figure}[h!]
	\includegraphics[width=8.7cm]{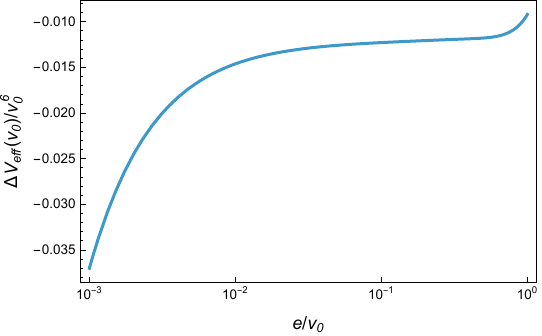}
\caption{
Shifted renormalized effective potential evaluated at the broken
minimum, $\Delta V_{\rm eff}^{\rm ren}(v_0)$, normalized to $v_0^6$,
as a function of the normalized gauge coupling $e/v_0$. The
parameters are fixed to $\lambda/v_0^2=0.05$,
$gv_0^2=1/50$, and $\mu/v_0^2=7\times10^{-1}$. The negative values
of $\Delta V_{\rm eff}^{\rm ren}(v_0)$ indicate that the broken phase
remains energetically favored throughout the range shown.
}
	\label{Fig2}
\end{figure}
A more quantitative view of the gauge-coupling dependence is provided
by Fig.~\ref{Fig2}, which displays the shifted renormalized effective
potential evaluated at the broken minimum,
$\Delta V_{\rm eff}^{\rm ren}(v_0)$, normalized to $v_0^6$, as a
function of the normalized gauge coupling $e/v_0$. The
parameters are fixed to 
$\lambda/v_0^2=0.05$, $gv_0^2=1/50$, and
$\mu/v_0^2=7\times10^{-1}$. As $e/v_0$ increases, the vacuum energy
rises monotonically toward less negative values, indicating that
gauge-field fluctuations progressively reduce the depth of the broken
minimum. The effect becomes more pronounced for
$e/v_0\sim1$, where the dependence on the gauge coupling
departs from its nearly linear behavior. Nevertheless, the vacuum
energy remains negative throughout the parameter range shown,
confirming that the spontaneously broken phase continues to be the
energetically preferred configuration.

\begin{figure}[h!]
	\includegraphics[width=8.7cm]{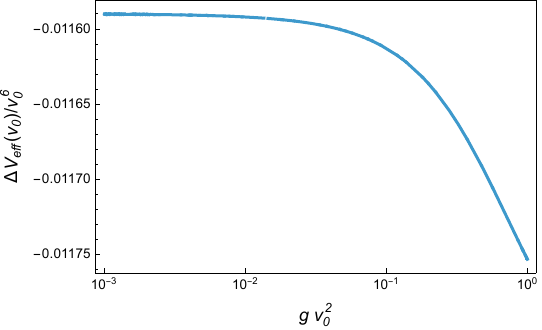}
\caption{
Shifted renormalized effective potential evaluated at the broken
minimum, $\Delta V_{\rm eff}^{\rm ren}(v_0)$, normalized to $v_0^6$, as
a function of the dimensionless higher-derivative coupling $gv_0^2$.
The parameters are fixed to $\lambda/v_0^2=0.05$, $e/v_0=4\times10^{-1}$, and
$\mu/v_0^2=2\times10^{-1}$. As the higher-derivative coupling
increases, the vacuum energy decreases monotonically, indicating a
slight enhancement of the stability of the spontaneously broken phase.
}
	\label{Fig3}
\end{figure}
To isolate the contribution of the higher-derivative CS
interaction, we fix the gauge coupling and vary the dimensionless
coupling $gv_0^2$. The resulting behavior is presented in
Fig.~\ref{Fig3}, which shows the shifted renormalized effective
potential evaluated at the broken minimum,
$\Delta V_{\rm eff}^{\rm ren}(v_0)$, normalized to $v_0^6$, as a
function of $gv_0^2$. The remaining parameters are fixed to
 $\lambda/v_0^2=0.05$, $e/v_0=4\times10^{-1}$, and
$\mu/v_0^2=2\times10^{-1}$. Unlike the gauge coupling, whose quantum
fluctuations tend to reduce the depth of the broken minimum, increasing
the higher-derivative coupling produces a smooth monotonic decrease of
the vacuum energy. Although the effect is quantitatively modest over
the parameter range considered, it indicates that the higher-derivative
interaction enhances the energetic preference for the spontaneously
broken phase.

\begin{figure}[h!]
	\includegraphics[width=8.7cm]{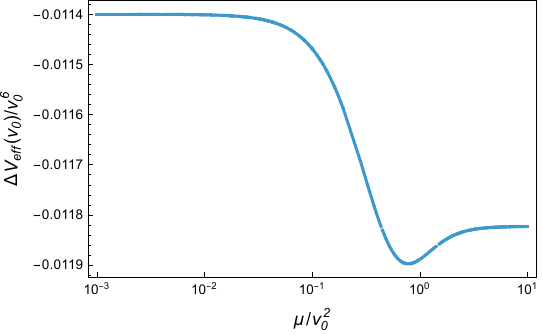}
\caption{
Shifted renormalized effective potential evaluated at the broken
minimum, $\Delta V_{\rm eff}^{\rm ren}(v_0)$, normalized to $v_0^6$,
as a function of the dimensionless CS coefficient
$\mu/v_0^2$. The parameters are fixed to 
$\lambda/v_0^2=0.05$, $e/v_0=4\times10^{-1}$, and
$gv_0^2=1/50$. The vacuum energy exhibits a nonmonotonic dependence
on $\mu/v_0^2$, while remaining below the symmetric configuration
throughout the range shown.
}
	\label{Fig4}
\end{figure}
Finally, we examine the dependence of the vacuum energy on the
conventional Chern--Simons coefficient. The corresponding results are
presented in Fig.~\ref{Fig4}, which displays the shifted renormalized
effective potential evaluated at the broken minimum,
$\Delta V_{\rm eff}^{\rm ren}(v_0)$, normalized to $v_0^6$, as a
function of the dimensionless parameter $\mu/v_0^2$. The remaining
parameters are fixed to  
$\lambda/v_0^2=0.05$, $e/v_0=4\times10^{-1}$, and
$gv_0^2=1/50$. Unlike the monotonic dependence observed for the
higher-derivative coupling in Fig.~\ref{Fig3}, the vacuum energy
exhibits a nonmonotonic behavior as a function of
$\mu/v_0^2$. Starting from small values of the Chern--Simons
coefficient, the broken minimum becomes progressively deeper until it
reaches a minimum at intermediate values of $\mu/v_0^2$. For larger
values of the CS coefficient, the vacuum energy increases
slightly and approaches an approximately constant value, indicating
that its influence on the vacuum structure becomes progressively
weaker.

\section{Final remarks}\label{sec:conclusions}
In this work, we have investigated the origin of dipole ghost sectors 
in gauge theories and their relation to the multiplicity of classical 
solutions. Rather than introducing dipole ghosts through auxiliary fields 
satisfying fourth-order equations of motion, we developed a constructive 
framework in which these sectors emerge directly from the structure of the 
classical equations. To establish this connection, we first analyzed 
covariant Maxwell theory in $(2+1)$ dimensions, where all relevant 
ingredients can be identified explicitly. We then extended the analysis 
to the constrained higher-order CS theory and showed that the 
same mechanism remains operative in the presence of constraints.

Our results reveal a direct correspondence between repeated roots 
of the characteristic polynomial, generalized solutions of the form 
$t e^{-i\omega t}$, Jordan-block structures, mixed oscillator algebras, 
and the appearance of dipole ghost sectors in the quantum theory. 
In particular, we showed that the degenerate sector is not described 
by a collection of independent canonical oscillators, but instead gives rise to a nontrivial 
oscillator algebra containing both positive and negative-norm states. 
The resulting Feynman propagator reproduces the characteristic 
double-pole structure associated with dipole ghosts, establishing a 
direct link between the multiplicity of classical solutions and their 
quantum realization.

An important outcome of the analysis is that the dipole ghost does not 
appear as an independent ingredient of the theory, but rather emerges 
dynamically from the degeneracy of the classical equations of motion. 
From this perspective, the double pole of the propagator should be viewed 
as a consequence of the 
underlying multiplicity structure rather than as a fundamental starting 
point. This provides a unified interpretation of dipole ghost sectors 
in terms of generalized eigenmodes and non-diagonalizable dynamical operators.

We also investigated the effects of spontaneous symmetry breaking in a
higher-order CS model coupled to a charged scalar field. The
Higgs mechanism generates a gauge-boson mass that lifts the degeneracy
responsible for the dipole ghost sector, replacing it with a spectrum
of ordinary massive excitations. The corresponding renormalized
one-loop effective potential was computed, providing a dynamical probe
of the vacuum structure of the broken phase. Our analysis shows that
the broken vacuum remains energetically favored after the inclusion of
quantum corrections over the parameter range considered. Consequently,
the quantum theory dynamically selects a vacuum characterized by a
nondegenerate massive gauge spectrum, rather than the degenerate
spectrum associated with the dipole ghost sector. These results provide
evidence that the Higgs mechanism not only removes the spectral
degeneracy responsible for the dipole ghost, but also stabilizes the
corresponding ghost-free phase at the quantum level.

Another noteworthy result is that the Gupta-Bleuler prescription can be consistently implemented 
in both the Maxwell and the extended CS theories. Since the CS contribution 
is transverse, it does not modify the divergence of the equations of motion. Consequently, 
the Lorentz condition $\Lambda_{CS}=\partial_\mu A^{\mu,CS}$ satisfies a free wave equation and may be 
imposed weakly on the physical Hilbert space through $\Lambda^{(+)}_{CS} |\mathrm{phys}\rangle=0$. 
This shows that the emergence of dipole ghost sectors is compatible with the standard covariant 
quantization procedure and does not obstruct the definition of a physical subspace.

Beyond the specific models considered here, the framework developed 
in this work provides a systematic method for identifying and quantizing 
degenerate sectors in field theories. Since multiplicity of solutions and 
Jordan structures arise in a wide variety of physical contexts, the present 
approach may prove useful in the study of generalized gauge theories, 
higher-derivative models, critical theories, and other systems where 
non-diagonalizable dynamics plays an important role.

\section*{Acknowledgments}
 The research of CMR
was partially supported by Fondecyt Regular project No. 1241369; and wants 
to thank the kind hospitality at 
Universidad Nacional Autónoma de M\'exico (UNAM) where this work was finished. CR acknowledges support from the 
Universidad San Sebastián 2026 Postdoctoral Researchers Attraction Program, grant USS-FIN-26-PDOC-03. AS was supported by DGAPA-UNAM under Grant No. PAPIIT-IN116326.
\appendix
\section{The oscillator algebra}\label{App:A}
In this Appendix, we verify the oscillator algebra presented 
in Eqs.~\eqref{algebraoscilator}. For this purpose, we consider the following general algebra involving a set of a priori undetermined functions $N_i(k)$ with $i=1\dots 6$
\begin{subequations}
\begin{align}
    \big[a_{{k}},a_{{q}}^\dagger\big]&= (2\pi)^2 N_1(k) 
    \delta^{(2)}(\vec{k}-\vec{q})\,, \\
    \big[a_{{k}},b_{{q}}^\dagger\big]&= (2\pi)^2 N_2(k)
    \delta^{(2)}(\vec{k}-\vec{q}) \,, \notag  \\
    \big[a_{{k}},c_{{q}}^\dagger\big]&= (2\pi)^2 N_3(k)
    \delta^{(2)}(\vec{k}-\vec{q}) \,,\notag \\
    \big[b_{{k}},b_{{q}}^\dagger\big]&= (2\pi)^2 N_4(k)
    \delta^{(2)}(\vec{k}-\vec{q}) \,,\notag \\
    \big[b_{{k}},c_{{q}}^\dagger\big]&= (2\pi)^2 N_5(k)
    \delta^{(2)}(\vec{k}-\vec{q})  \,,\notag  \\
    \big[c_{{k}},c_{{q}}^\dagger\big]&= (2\pi)^2 N_6(k)
    \delta^{(2)}(\vec{k}-\vec{q}) \,.\notag
\end{align} \label{ansatz-usual}
\end{subequations}
We impose the equal time commutation relations 
\begin{align}
\big[ A_\mu(x),\Pi^\nu(y)\big]&=i \delta^\nu_\mu \delta^{(2)}
(\vec{x}-\vec{y})\,, \label{Can-Usual}
\end{align}
and the vanishing ones
  \begin{align}
  \big[ A_\mu(x),A_\nu(y)\big]&=0 \label{Can-Usual-A}\,,\\
    \big[ \Pi^\mu(x),\Pi^\nu(y)\big]&=0  \label{Can-Usual-P}\,.
\end{align}
The canonical momentum conjugated to $A_\mu$ field is given by
\begin{align}
    \Pi^\mu:=- F^{0\mu}  -\frac{1}{\xi} \big(\partial\cdot A\big)\eta^{0\mu} \,.
\end{align}
We define the momenta associated to each degree of freedom as
\begin{align}
    \Pi^\mu_{(I)}= -\Big(\partial^0 A_{(I)}^\mu - \partial^\mu A_{(I)}^0\Big)-\frac{1}{\xi}\Big(\partial \cdot A_{(I)}\Big) \eta^{0\mu} \,,
\end{align}
with $  I=\lbrace \text{G,D,T} \rbrace $. Explicitly, they are
\begin{align}
    \Pi_{\text{G}}^\mu&= 0 \label{Pi-G}\,,\\
    \Pi_{\text{D}}^\mu&=  \bigg(\frac{2}{1-\xi}  \bigg) \int \frac{\mathrm{d}^2k}{(2\pi)^2}
    \big( -k^\mu+2   \vert\vec{k}\vert  n^\mu\big)\notag  \\ &\hspace{6em}
    \times \big( b_{{k}} e^{-i k\cdot x}+    b^\dagger_{{k}} e^{ik\cdot x}  \big)_{k_0=\omega }  \,, \\
    \Pi_{\text{T}}^\mu&=  \int \frac{\mathrm{d}^2k}{(2\pi)^2}i\vert\vec{k}\vert
    \epsilon^{\mu\alpha  \beta} k_\alpha n_\beta \big(  c_{{k}} e^{-i k\cdot x}-  c^\dagger_{{k}} e^{ik\cdot x}\big)_{k_0=\omega}\,,
\end{align}
\subsection{The commutator $\big[\Pi^\mu(x),\Pi^\nu(y)\big]$}
Due to Eq.~\eqref{Pi-G}, the only contributions to this commutator are
\begin{align}
    \big[\Pi^\mu(x),\Pi^\nu(y)\big]=&\big[\Pi_{\text{D}}^\mu(x), \Pi_{\text{D}}^\nu(y)\big]+\big[ \Pi_{\text{D}}^\mu(x), \Pi_{\text{T}}^\nu(y)\big]\notag \\
    &+\big[ \Pi_{\text{T}}^\mu(x), \Pi_{\text{D}}^\nu(y)\big]  +\big[ \Pi_{\text{T}}^\mu(x), \Pi_{\text{T}}^\nu(y)\big] \,.
\end{align}
Using Eqs.~\eqref{ansatz-usual}, each contribution is given by
\begin{align}
    \big[\Pi_{\text{D}}^\mu(x), \Pi_{\text{D}}^\nu(y)\big] 
     =&\bigg(\frac{2}{1-\xi}  \bigg)^2 \int \frac{\mathrm{d}^2k}{(2\pi)^2}(2\vert\vec{k}\vert) N_4(k) \notag \\
    &\times\big[    k^\mu n^\nu   +   n^\mu k^\nu-2   \vert\vec{k}\vert   n^\mu      n^\nu\big]   e^{ik\cdot( x-  y)}  \,,
\end{align}
\begin{align}
    \big[ \Pi_{\text{D}}^\mu(x), \Pi_{\text{T}}^\nu(y)\big]=&\bigg(\frac{2}{1-\xi}  \bigg) \int \frac{\mathrm{d}^2k}{(2\pi)^2} (2i\vert\vec{k}\vert)N_5(k) \notag \\
    &\times \big[k^\mu -n^\mu \vert\vec{k}\vert \big]\big[\epsilon^{\nu \alpha  \beta} k_\alpha n_\beta  \big]     e^{ik\cdot( x-  y)}   \,,
\end{align}
\begin{align}
    \big[\Pi_{\text{T}}^\mu(x),\Pi_{\text{D}}^\nu(y)\big]=& \bigg(\frac{2}{1-\xi}  \bigg)\int \frac{\mathrm{d}^2k}{(2\pi)^2}(-2i\vert\vec{k}\vert) N_5(k)\notag \\
    &\times \big[\epsilon^{\mu\alpha  \beta} k_\alpha n_\beta \big]\big[  k^\nu-   \vert\vec{k}\vert  n^\nu\big] e^{ik\cdot( x- y)} \,, 
\end{align}
and
\begin{align}
    \big[\Pi_{\text{T}}^\mu(x),\Pi_{\text{T}}^\nu(y)\big] =& \int \frac{\mathrm{d}^2k}{(2\pi)^2}(i\vert\vec{k}\vert \vert\vec{q}\vert )N_6(k)\big[\epsilon^{\mu\alpha  \beta} k_\alpha n_\beta \big]\notag \\
    &\times \big[\epsilon^{\nu\rho  \sigma} k_\rho n_\sigma \big] \big(     e^{-i k\cdot( x-  y)}- e^{ik\cdot (x-  y)}   \big) \notag \\
    &=   0 \,.
\end{align}
Inserting the previous commutators results in
\begin{align}
    &\big[\Pi^\mu(x),\Pi^\nu(y)\big]\notag \\
    =&\bigg(\frac{2}{1-\xi}  \bigg)  \int \frac{\mathrm{d}^2k}{(2\pi)^2}(2\vert\vec{k}\vert) \bigg[N_4(k)\bigg(\frac{2}{1-\xi}  \bigg)\big(    k^\mu n^\nu   +   n^\mu k^\nu\notag \\
    &-2   \vert\vec{k}\vert   n^\mu      n^\nu\big) +iN_5(k)\vert\vec{k}\vert^2\epsilon^{\mu\nu\lambda}      n_\lambda \bigg]e^{ik\cdot( x-  y)}  \,,
\end{align}
According to Eq.~\eqref{Can-Usual-P}, this leads to
\begin{eqnarray}
   N_4(k)     &=&0 \label{N4}\,, \\
            N_5(k)  &=&0 \label{N5}\,.
\end{eqnarray}
Consequently, the 
$\text{D}-\text{D}$, $\text{D}-\text{T}$, and $\text{T}-\text{D}$ commutators do not contribute.
\subsection{The commutator $\big[A_\mu(x),\Pi^\nu(y)\big]$}
Considering Eqs.~\eqref{N4},~\eqref{N5}, the canonical commutator Eq.~\eqref{Can-Usual} receives contributions only from
\begin{align}
    \big[A_\mu(x),\Pi^\nu(y)\big]=&\big[A^{\text{G}}_\mu(x),\Pi_{\text{D}}^\nu(y)\big]+\big[A^{\text{G}}_\mu(x),\Pi_{\text{T}}^\nu(y)\big]\notag \\
    &+\big[A^{\text{T}}_\mu(x),\Pi_{\text{T}}^\nu(y)\big] \,.
\end{align}
Explicitly, the contributions read
\begin{align}
    \big[A^{\text{G}}_\mu(x),\Pi_{\text{D}}^\nu(y)\big] =&  \bigg(\frac{2}{1-\xi}  \bigg)\int \frac{\mathrm{d}^2k}{(2\pi)^2} (-2i)N_2(k)\notag \\
    &\times \big[\vert\vec{k}\vert(n_\mu k^\nu+k_\mu n^\nu ) -k_\mu k^\nu  \big]e^{ik\cdot( x-  y)} \,,
\end{align}
\begin{align}
    \big[ A_\mu^{\text{G}}(x),\Pi^\nu_{\text{T}}(y)\big]=&\int \frac{\mathrm{d}^2k}{(2\pi)^2}(2 \vert\vec{k}\vert ^2) N_3(k)    \notag \\
    &\times n_\mu\big[\epsilon^{\nu \alpha\beta} k_\alpha  n_\beta \big]e^{ik\cdot( x-  y)} \,,
\end{align}
\begin{align}
   \big[ A_\mu^{\text{T}}(x),\Pi^\nu_{\text{T}}(y)\big]=&  \int \frac{\mathrm{d}^2k}{(2\pi)^2}(-2i\vert\vec{k}\vert) N_6(k) \big[-\vert\vec{k}\vert^2\delta_\mu^\nu  \notag \\
   & -k_\mu k^\nu+\vert\vec{k}\vert  (k_\mu n^\nu+n_\mu  k^\nu )\big]   e^{ik\cdot( x-  y)} \,.  
\end{align}
Replacing in the canonical commutator we obtain
\begin{align}
    &\big[A_\mu(x),\Pi^\nu(y)\big] \notag \\
    =&\int \frac{\mathrm{d}^2k}{(2\pi)^2}(-2i)\bigg[  \bigg(\frac{2}{1-\xi}  \bigg)N_2(k)\big(\vert\vec{k}\vert(n_\mu k^\nu+k_\mu n^\nu )\notag \\
    &-k_\mu k^\nu  \big)+i N_3(k)     \vert\vec{k}\vert ^2  n_\mu\big[\epsilon^{\nu \alpha\beta} k_\alpha  n_\beta \big]+N_6(k)  \vert\vec{k}\vert\notag \\
    & \times \big(-\vert\vec{k}\vert^2\delta_\mu^\nu-k_\mu k^\nu +\vert\vec{k}\vert  (k_\mu n^\nu+n_\mu  k^\nu )\big)\bigg]e^{ik\cdot( x-  y)}  \,,
\end{align}
According to Eq.~\eqref{Can-Usual}, this leads to
\begin{eqnarray}
    N_2(k)&=&-\frac{(1-\xi)}{4\vert\vec{k}\vert^2}  \,, \\
   N_3(k)&=&0 \label{N3}\,,\\
   N_6(k)&=&  \frac{1}{2 \vert\vec{k}\vert^3   } \,.
\end{eqnarray}
The result Eq.~\eqref{N3} implies that there is no contribution from the sectors $\text{G}-\text{T}$ and $\text{T}-\text{G}$.
\subsection{The commutator $\big[A_\mu(x),A_\nu(y)\big]$}
Finally, considering~\eqref{N3},~\eqref{N4},~\eqref{N5}, the commutator Eq.~\eqref{Can-Usual-A} receives contributions only from the following sectors:
\begin{align}
\big[A_\mu(x),A_\nu(y)\big]=&\big[A^{\text{G}}_\mu(x),
A^{\text{G}}_\nu(y)\big]+\big[A^{\text{G}}_\mu(x),A^{\text{D}}_\nu(y)\big] \notag \\
    &+\big[A^{\text{D}}_\mu(x),A^{\text{G}}_\nu(y)\big]+\big[A^{\text{T}}_\mu(x),A^{\text{T}}_\nu(y)\big] \,.
\end{align}
The explicit contributions are given by
\begin{align}
 \big[A^{\text{G}}_\mu(x),A^{\text{G}}_\nu(y)\big]=&\int \frac{\mathrm{d}^2k}{(2\pi)^2}N_1(k) (-2\vert\vec{k}\vert) \notag \\
 &\times \big(n_\mu  k_\nu  +k_\mu  n_\nu - 2\vert\vec{k}\vert n_\mu  n_\nu \big)e^{ik\cdot(x-y)} \,,
\end{align}
\begin{align}
    \big[A^{\text{G}}_\mu(x),A^{\text{D}}_\nu(y)\big]=&(1-\xi)\int \frac{\mathrm{d}^2k}{(2\pi)^2}   \frac{2i}{4\vert\vec{k}\vert^2}  \Big[   (  k_\mu  k_\nu \notag \\
    &-\vert\vec{k}\vert(k_\mu  n_\nu   +n_\mu    k_\nu)   +2\vert\vec{k}\vert^2 n_\mu n_\nu  )  t   \notag \\
    &-i\bigg(\frac{1+\xi}{1-\xi} \bigg)(  k_\mu  -\vert\vec{k}\vert n_\mu  )  n_\nu \Big]e^{ik\cdot( x-  y)} \,,
\end{align}
\begin{align}
\big[A^{\text{D}}_\mu(x),A^{\text{G}}_\nu(y)\big]=& -(1-\xi) \int \frac{\mathrm{d}^2k}{(2\pi)^2}\frac{2i}{4\vert\vec{k}\vert^2} \Big[ \big( k_\mu  k_\nu \notag \\
&-\vert\vec{k}\vert (k_\mu  n_\nu +n_\mu  k_\nu  )+2\vert\vec{k}\vert^2 n_\mu n_\nu     \big)t\notag \\
      &+i\bigg(\frac{1+\xi}{1-\xi} \bigg) n_\mu(k_\nu-\vert\vec{k}\vert n_\nu )     \Big]e^{ik\cdot( x- y)}   \,,
\end{align}  
and 
\begin{align}
\big[A^{\text{T}}_\mu(x),A^{\text{T}}_\nu(y)\big]=& \int \frac{\mathrm{d}^2k}{(2\pi)^2}\frac{1}{2\vert\vec{k}\vert^3}\big[\epsilon_{\mu\alpha\beta  }k^\alpha n^\beta\big]\big[\epsilon_{\nu\rho\sigma  }k^\rho n^\sigma\big]\notag \\
&\times \big(   e^{-i k\cdot( x-  y)}-e^{ik\cdot( x-  y)} \big) \notag \\
&\equiv  0 \,.
\end{align}
Combining these results, we obtain
 \begin{align}
     \big[A_\mu(x),A_\nu(y)\big]=&\int \frac{\mathrm{d}^2k}{(2\pi)^2}   \bigg[-2 N_1(k)  \vert\vec{k}\vert  +\frac{(1+\xi)}{2\vert\vec{k}\vert^2}  \bigg]\notag \\
     &\times \big[    n_\mu k_\nu +     k_\mu n_\nu-2\vert\vec{k}\vert n_\mu n_\nu      \big]e^{ik\cdot( x-  y)}  \,,
 \end{align}
which determines the remaining function,
 \begin{eqnarray}
      N_1(k)&=& \frac{(1+\xi)}{4\vert\vec{k}\vert^3}      \,. 
 \end{eqnarray}
We therefore arrive at the algebra~\eqref{algebraoscilator}
consistent with the canonical commutation relations.
\section{The propagator in $2+1$ Maxwell theory}\label{App:B}
The Feynman propagator is defined as vacuum expectation value of the time-ordered product of the field operators at different spacetime points:
\begin{align}
    D^F_{\mu\nu}(x-y)&= \theta(x^0-y^0)\langle 0\vert A_\mu(x)A_\nu(y)\vert 0 \rangle \notag \\
    &\hspace{1em}+\theta(y^0-x^0)\langle 0\vert A_\nu(y) A_\mu(x)\vert 0 \rangle \,,
\end{align}
with $\theta(x)$ being the Heaviside step function. 

Using the commutation relations~\eqref{algebraoscilator}
, only four non-vanishing expectation value 
contribute to each time ordering. Therefore,
\begin{align}
 & \langle 0\vert A_\mu(x)A_\nu(y)\vert 0 \rangle  \notag \\
 \hspace{0.5em}& =\langle 0\vert  A^{\text{G}}_\mu(x) A^{\text{G}}_\nu(y) \vert 0 \rangle +\langle 0\vert  A^{\text{G}}_\mu(x) A^{\text{D}}_\nu(y) \vert 0 \rangle \notag \\
   &\hspace{1em}+\langle 0\vert  A^{\text{D}}_\mu(x) A^{\text{G}}_\nu(y) \vert 0 \rangle +\langle 0\vert  A^{\text{T}}_\mu(x) A^{\text{T}}_\nu(y) \vert 0 \rangle \,,  
\end{align}
and
\begin{align}
    &\langle 0\vert A_\nu(y) A_\mu(x)\vert 0 \rangle \notag \\
    \hspace{0.5em}& = \langle 0\vert  A^{\text{G}}_\nu(y) A^{\text{G}}_\mu(x) \vert 0 \rangle +\langle 0\vert  A^{\text{G}}_\nu(y) A^{\text{D}}_\mu(x) \vert 0 \rangle \notag \\
   &\hspace{1em}+\langle 0\vert  A^{\text{D}}_\nu(y) A^{\text{G}}_\mu(x) \vert 0 \rangle +\langle 0\vert  A^{\text{T}}_\nu(y) A^{\text{T}}_\mu(x) \vert 0 \rangle\,. 
\end{align}
Evaluating each term separately yields
\begin{align}
    \langle 0\vert  A^\text{G}_\mu(x) A^\text{G}_\nu(y) \vert 0 \rangle&= \int \frac{\mathrm{d}^2k}{(2\pi)^2}  \frac{(1+\xi) }{4\omega^3}k_\mu k_\nu     e^{-i k\cdot( x-  y)} \,, 
\end{align}
\begin{align}
    \langle 0\vert  A^{\text{G}}_\mu(x) A^{\text{D}}_\nu(y) \vert 0 \rangle&= i  \int \frac{\mathrm{d}^2k}{(2\pi)^2} \frac{(1-\xi) }{4\omega^2}    \notag \\
    &\hspace{3em} \times k_\mu  v_\nu^*(y_0,k)   e^{-i k\cdot( x-  y)} \,, 
\end{align}
\begin{align}
     \langle 0\vert  A^{\text{D}}_\mu(x) A^{\text{G}}_\nu(y) \vert 0 \rangle&=     -i   \int \frac{\mathrm{d}^2k}{(2\pi)^2}\frac{(1-\xi) }{4\omega^2}   \notag \\
     &\hspace{3em}\times v_\mu(x_0,k)  k_\nu  e^{-i k\cdot( x-  y)} \,, 
\end{align}
\begin{align}
    \langle 0\vert  A^{\text{T}}_\mu(x) A^{\text{T}}_\nu(y) \vert 0 \rangle&=\int \frac{\mathrm{d}^2k}{(2\pi)^2} \frac{1}{2 \omega^3   } \big[-\omega^2\eta_{\mu\nu}  -k_{\mu} k_{\nu}\notag \\
    &\hspace{1em}+\omega(k_{\mu}  n_{\nu} +n_{\mu } k_{ \nu})\big]      e^{-i k\cdot( x- y)} \,,
\end{align}
where the remaining contributions can be obtained performing $\mu \leftrightarrow \nu$ and $x \leftrightarrow y$. 

Adding all terms, one finds
\begin{align}
  & \langle 0\vert A_\mu(x)A_\nu(y)\vert 0\rangle \notag \\
   =&  \int \frac{\mathrm{d}^2k}{(2\pi)^2}  \bigg[\frac{(1+\xi) }{4\omega^3}  k_\mu k_\nu  +    i\frac{(1-\xi)}{4\omega^2}  \big[ k_\mu  v_\nu^*(y_0,k) \notag \\
   & -v_\mu(x_0,k)  k_\nu \big] +  \frac{1}{2 \omega^3   } \big[-\omega^2\eta_{\mu\nu}   -k_{\mu} k_{\nu} \notag \\
   &+\omega (k_{\mu}  n_{\nu} +n_{\mu } k_{ \nu})\big]  \bigg]      e^{-i k\cdot( x- y)} \,,
\end{align}
\begin{align}
  & \langle 0\vert A_\nu(y)A_\mu(x)\vert 0\rangle \notag \\
   =&  \int \frac{\mathrm{d}^2k}{(2\pi)^2}  \bigg[\frac{(1+\xi) }{4\omega^3}  k_\mu k_\nu  +    i\frac{(1-\xi)}{4\omega^2}  \big[ k_\nu  v_\mu^*(x_0,k) \notag \\
   & -v_\nu(y_0,k)  k_\mu \big] +  \frac{1}{2 \omega^3   } \big[-\omega^2\eta_{\mu\nu}   -k_{\mu} k_{\nu} \notag \\
   &+\omega (k_{\mu}  n_{\nu} +n_{\mu } k_{ \nu})\big]  \bigg]      e^{-i k\cdot( y- x)} \,.
\end{align}
Using the definition of $v_\mu$ given in Eq.~\eqref{v-def}, one can show that
\begin{align}
     & k_\mu  v_\nu^*(y_0,k) -v_\mu(x_0,k)  k_\nu  \notag \\
     &= -k_\mu  k_\nu  (x_0 -y_0) +i\bigg(\frac{1+\xi}{1-\xi} \bigg) (k_\mu n_\nu+n_\mu k_\nu) \,, \label{v-identity1}
\end{align}
and
\begin{align}
      &k_\nu  v_\mu^*(x_0,k) -v_\nu(y_0,k)  k_\mu \notag \\
       &=  -k_\mu  k_\nu  (y_0 -x_0) +i\bigg(\frac{1+\xi}{1-\xi} \bigg) (k_\mu n_\nu+n_\mu k_\nu) \,. \label{v-identity2}
\end{align}
Introducing the coordinate $z=x-y$, and using the previous identities allow us to write Feynman propagator as
\begin{align}
    D_{\mu\nu}^F(z)&= \theta(z^0)  \int \frac{\mathrm{d}^2k}{(2\pi)^2}  \bigg[- \frac{\eta_{\mu\nu}}{2 \omega   }   +     (1-\xi)   \bigg(\frac{1}{4\omega^2}\big( k_\mu n_\nu\notag \\
    &\hspace{1em}+n_\mu k_\nu- i  z^0 k_\mu k_\nu\big)  -\frac{ k_\mu k_\nu}{4\omega^3}\bigg)\bigg] e^{-i k\cdot z} \notag \\
    &\hspace{1em}+\theta(-z^0) \int \frac{\mathrm{d}^2k}{(2\pi)^2}  \bigg[-\frac{\eta_{\mu\nu} }{2 \omega    }   +  (1-\xi)     \bigg(\frac{1 }{4\omega^2} \big(k_\mu n_\nu\notag \\
    &\hspace{1em}+n_\mu k_\nu+i  z^0 k_\mu k_\nu\big)  -\frac{k_\mu k_\nu }{4\omega^3} \Big)\bigg]e^{i k\cdot z} \,.\label{Feynman-Usual-1}
\end{align}
In the terms proportional to $\theta(-z^0)$ we perform the change of variables $\vec{k}\rightarrow -\vec{k}$. By doing so, we rewrite Eq.~\eqref{Feynman-Usual-1} as
\begin{widetext}

\begin{align}
    D_{\mu\nu}^F(z)&= \theta(z^0)  \int \frac{\mathrm{d}^2k}{(2\pi)^2}    \bigg[- \frac{\eta_{\mu\nu}}{2\omega}  +      (1-\xi)   \bigg(\frac{k_\mu n_\nu+n_\mu k_\nu-i  z^0 k_\mu k_\nu}{(2\omega)^2}    +  \frac{(-2)k_\mu k_\nu}{(2\omega)^3}\bigg) \bigg]_{k_0=\omega}e^{-i\omega z^0} e^{i \vec{k}\cdot \vec{z}}   \notag \\
    &\hspace{0.5em}- \theta(-z^0) \int \frac{\mathrm{d}^2k}{(2\pi)^2}   \bigg[-\frac{\eta_{\mu\nu}}{(-2\omega)}     +    (1-\xi) \bigg(\frac{k_\mu n_\nu+n_\mu k_\nu-i  z^0 k_\mu k_\nu}{(-2\omega)^2}      +\frac{(-2)k_\mu k_\nu}{(-2\omega)^3}\bigg) \bigg]_{k_0=-\omega}e^{i\omega z^0}e^{i \vec{k}\cdot \vec{z}} \,.
\end{align}
\end{widetext}
The combination appearing in the integrand satisfies
\begin{align}
   \frac{\partial }{\partial k_0}\bigg[\frac{k_\mu  k_\nu}{(k_0\pm \vert\vec{k}\vert )^2}e^{- ik_0 z^0}\bigg]=& \bigg[\frac{k_\mu n_\nu+n_\mu k_\nu- i z^0 k_\mu k_\nu}{(k_0\pm\vert\vec{k}\vert )^2}\notag \\
   &+\frac{(-2)k_\mu  k_\nu}{(k_0\pm \vert\vec{k}\vert )^3}  \bigg]e^{-ik_0 z^0} \,,  \label{pole-identity}
\end{align}
Therefore, the integrand can be expressed in a form suitable for applying residue theorem. By using Eq.~\eqref{pole-identity}, the Feynman propagator becomes
\begin{align}
    D_{\mu\nu}^F(z)&= \theta(z^0)  \int \frac{\mathrm{d}^2k}{(2\pi)^2}    \bigg(- \frac{\eta_{\mu\nu}}{k_0+\omega}e^{-ik\cdot z}   \\
     &\hspace{1em}+(1-\xi)   \frac{\partial}{\partial k_0}\bigg[\frac{k_\mu k_\nu}{(k_0+\omega)^2}e^{-ik\cdot z}\bigg]\bigg)_{k_0=\omega}    \notag \\
    &\hspace{1em}-\theta(-z^0) \int \frac{\mathrm{d}^2k}{(2\pi)^2}  \bigg(- \frac{\eta_{\mu\nu}}{(k_0-\omega)}e^{-ik\cdot z}  \notag \\
    &\hspace{1em}+(1-\xi)   \frac{\partial}{\partial k_0}\bigg[\frac{k_\mu k_\nu}{(k_0-\omega)^2}e^{-ik\cdot z}\bigg]\bigg)_{k_0=-\omega} \notag   \,.   
\end{align}
Recalling the residue formula for a second-order pole,
\begin{eqnarray}
    \frac{1}{2\pi i}\oint_\gamma \mathrm{d}z\, \frac{F(z)}{(z-z_0)^2}= \lim_{z\rightarrow z_0} \frac{1}{(2-1)!}\frac{\mathrm{d}\,}{\mathrm{d}z} F(z) \,,
\end{eqnarray}
the contributions evaluated at  $k_0=\pm \omega$ are identified with the residue of the propagator $G_{\mu\nu}$ at those poles:
\begin{align}
    D_{\mu\nu}^F(z)&=  \int \frac{\mathrm{d}^2k}{(2\pi)^2}    \Big(\theta(z^0)\,\text{Res}\Big[iG_{\mu\nu}e^{-ik\cdot z},k_0=\omega\Big]    \notag \\
    &\hspace{1em}- \theta(-z^0)\,     \text{Res}\Big[iG_{\mu\nu}e^{-ik\cdot z},k_0=-\omega\Big]\Big)   \,,
\end{align}
with $G_{\mu\nu}$ given in Eq.~\eqref{Usual-Propagator}. 

For $z^0>0$, the imaginary part of $k_0$ requires the contour integration to be closed in the lower half-plane in clockwise direction. Consequently, the Feynman propagator becomes
\begin{align}
     D_{\mu\nu}^F(z)= - \int \frac{\mathrm{d}^2k}{(2\pi)^2}   \int_{\mathcal{C}_-}  \frac{\mathrm{d}k_0}{(2\pi)}G_{\mu\nu}e^{-ik\cdot z}\,.
\end{align}
If $z^0<0$, the imaginary part of $k_0$ must instead be closed in the upper half-plane in counter-clockwise direction. In this case we obtain
\begin{eqnarray}
    D_{\mu\nu}^F(z)= - \int \frac{\mathrm{d}^2k}{(2\pi)^2}    \int_{\mathcal{C}_+} \frac{\mathrm{d}k_0}{(2\pi)}G_{\mu\nu}e^{-ik\cdot z} \,.
\end{eqnarray}
Finally, the Feynman propagator, with the corresponding Feynman prescription, is given by 
\begin{eqnarray}
    D^F_{\mu\nu}(\xi,k)=- \frac{ \eta_{\mu\nu}}{k^2+i\epsilon } +(1-\xi)  \frac{k_\mu  k_\nu  }{(k^2+i\epsilon )^2} \,.
\end{eqnarray}
\section{Faddeev-Popov operator in the broken phase}\label{app:FP}
In this Appendix we derive the Faddeev-Popov operator associated with
the $R_\xi$ gauge employed in Sec.~\ref{SSBIV}. As discussed in the main text,
both the Maxwell and the extended CS terms are exactly gauge
invariant. Consequently, the Faddeev-Popov operator is completely
determined by the gauge-fixing functional and is therefore independent
of the higher-derivative CS coupling $g$. We present its
explicit derivation here for completeness.

The gauge-fixing function is
\begin{align}
F[A,\chi]=\partial_\mu A^\mu-\xi ev\chi \,.
\end{align}
Under an infinitesimal Abelian gauge transformation,
\begin{align}
\delta A_\mu=\partial_\mu \alpha\,,
\end{align}
while the scalar field transforms as
\begin{align}
\delta\Phi=ie\alpha\Phi \,.
\end{align}
Using
\begin{align}
\Phi=\frac{1}{\sqrt2}\left(v+h+i\chi\right)\,,
\end{align}
one obtains
\begin{align}
\delta h&=-e\alpha\chi,\\\delta\chi&=e\alpha(v+h)\,.
\end{align}

Therefore,
\begin{align}
\delta F&=\partial_\mu\delta A^\mu-\xi ev\,\delta\chi\,, \nonumber\\
&=\Box\alpha-\xi e^2v(v+h)\alpha \,.
\end{align}
Thus, the Faddeev-Popov operator is
\begin{align}
\mathcal{M}_{FP}=\frac{\delta F}{\delta\alpha}=\Box-\xi e^2v(v+h)\,.
\end{align}
At quadratic order in the fluctuations, only the background-dependent
part contributes, and hence
\begin{align}
\mathcal{M}^{(2)}_{FP}=\Box-\xi e^2v^2=\Box-\xi m_\gamma^2(v)\,.
\end{align}
Exponentiation of the Faddeev-Popov determinant introduces a pair of
Grassmann ghost fields,
\begin{align}
\det\mathcal{M}_{FP}=\int\mathcal{D}\bar 
c\,\mathcal{D}c\,\exp\left[i\int\mathrm{d}^3x\,\mathcal{L}_{\rm FP}
\right]\,,
\end{align}
with the ghost Lagrangian
\begin{align}
\mathcal{L}_{\rm FP}=\bar c\left(-\Box+\xi m_\gamma^2(v)\right)c\,.
\label{GhostLagrangianAppendix}
\end{align}

Thus, although the extended CS term modifies the physical
gauge-field spectrum through the quadratic operator discussed in
Sec.~\ref{SSBIV}, it does not alter the Faddeev-Popov operator, which depends
exclusively on the gauge-fixing functional. Consequently, the ghost
sector is identical to that of the conventional Abelian Higgs model,
with the gauge-dependent mass
\begin{align}
m_c^2(v)=\xi m_\gamma^2(v)\,.
\end{align}


 \end{document}